\documentclass[format=acmlarge, anonymous=false, screen=true]{acmart}
\renewcommand\footnotetextcopyrightpermission[1]{}
\newcommand{\myfigsize}{0.7} 

\newcommand{\code}[1]{\mbox{\texttt{#1}}}
\newcommand{\var}[1]{\mbox{\texttt{#1}}}
\newcommand{\bit}[1]{\mbox{\texttt{#1}}}

\usepackage{xcolor, soul}

\usepackage{pbox}
\usepackage{makecell}
\usepackage{tikz}
\usetikzlibrary{calc}

\newcommand{\method}[1]{\textup{\textsf{#1}}}
\newcommand{\Sum}{\code{sum}}
\newcommand{\Update}{\code{update}}
\newcommand{\Access}{\code{access}}
\newcommand{\Search}{\code{search}}

\newcommand{\fentree}{\method{Fenwick-Tree}}
\newcommand{\segtree}{\method{Segment-Tree}}

\newcommand{\sst}{\method{ST}}
\newcommand{\ft}{\method{FT}}

\usepackage{graphicx}
\usepackage{balance}
\usepackage{xcolor}
\usepackage{url}
\usepackage{amsmath}
\usepackage{multirow}
\usepackage{booktabs}
\usepackage{makecell}
\usepackage[caption=false]{subfig}
\usepackage{wrapfig}
\usepackage{xcolor, soul}
\usepackage{siunitx}
\usepackage{enumitem}
\usepackage{hyperref}
\usepackage{rotating}
\usepackage{tablefootnote}
\usepackage{listings}
\usepackage{afterpage}

\setlist[itemize,1]{leftmargin=5mm,itemsep=0mm}
\setlist[enumerate,1]{leftmargin=5mm,itemsep=0mm}

\definecolor{keywords}{HTML}{dd1c77}
\definecolor{strings}{HTML}{00999A}
\definecolor{comments}{HTML}{31a354}
\definecolor{identifiers}{HTML}{000000}
\newcommand{\results}{results_and}

\begin{document}
\title{Practical Trade-Offs for the Prefix-Sum Problem}

%

\author{Giulio Ermanno Pibiri}
\affiliation{%
  \institution{ISTI-CNR} 
}
\email{giulio.ermanno.pibiri@isti.cnr.it}

\author{Rossano Venturini}
\affiliation{%
  \institution{University of Pisa} 
}
\email{rossano.venturini@unipi.it}
\begin{abstract}
Given an integer array $A$, the \emph{prefix-sum problem} is 
to answer $\Sum(i)$ queries that return the sum of the elements
in $A[0..i]$, knowing that the integers in $A$ can be changed.
It is a classic problem in data structure design with a wide range of applications
in computing from coding to databases.

In this work,
we propose and compare several and practical solutions to this problem,
showing that new trade-offs between the performance of queries and updates
can be achieved on modern hardware.
\end{abstract}
\maketitle
\renewcommand{\shortauthors}{G.\,E. Pibiri and R. Venturini}

\section{Introduction}\label{sec:introduction}


The \emph{prefix-sum problem} is defined as follows.
Given an array $A[0..n)$ of integers and an index $0 \leq i < n$,
we are asked to support the following three operations
as efficiently as possible.
\begin{itemize}
\item $\Sum(i)$ returns the quantity $\sum_{k=0}^i A[k]$.
\item $\Update(i,\Delta)$ sets $A[i]$ to $A[i]+\Delta$, where $\Delta$
is a quantity that fits in $\delta$ bits.
\item $\Access(i)$ returns $A[i]$.
\end{itemize}
(Note that $\Access(i)$ can be computed as $\Sum(i)-\Sum(i-1)$ for $i > 0$ and
$\Access(0) = \Sum(0)$. Therefore, we do not consider this operation in the following.
Also, a \emph{range-sum} query $\Sum(i,j)$, asking for the sum of the elements
in $A[i..j]$, is computed as $\Sum(j) - \Sum(i-1)$.)

It is an icon problem in data structure design
and has been studied rather extensively from a theoretical point of view~\cite{FS89,yao1985complexity,hampapuram1998optimal,dietz1989optimal,raman2001succinct,hon2011succinct,patrascu2006logarithmic,chan2010counting,BCGSVV15,bille2017succinct}
given its applicability to
many areas of computing, such as coding, databases, parallel
programming, dynamic data structures and others~\cite{blelloch1990pre}.
For example, one of the most notable practical applications of this problem
is for \emph{on-line analytical processing} (OLAP) in databases.
An OLAP system relies on the popular \emph{data cube} model~\cite{gray1997data},
where the data is represented as a $d$-dimensional array.
To answer an aggregate query on the data cube, a prefix-sum query
is formulated (see the book edited by~\citet{toth2017handbook} --
Chapter 40, \emph{Range Searching} by P.K. Agarwal).

\paragraph*{Scope of this Work and Overview of Solutions}
Despite the many theoretical results, that we review in Section~\ref{sec:related_work},
literature about the prefix-sum problem lacks a thorough
experimental investigation which is our concern with this work.
We aim at determining the fastest single-core solution
using \emph{commodity hardware and software},
that is,
a recent manufactured processor with commodity architectural features
(such as pipelined instructions,
branch prediction, cache hierarchy, and SIMD instructions~\cite{SIMDIntel}),
executing C++ compiled with a recent optimizing compiler.
(We do not take into account parallel algorithms here, nor solutions devised
for specialized/dedicated hardware that would limit the usability of our software.
We will better describe our experimental setup in Section~\ref{sec:setup}.)

As a warm-up, let us now consider two trivial solutions to the problem that
will help reasoning about the different trade-offs.
A first option is to leave $A$ as given.
It follows that queries are supported in
$O(n)$ by scanning the array and updates take $O(1)$ time.
Otherwise, we can pre-compute the result of $\Sum(i)$
and save it in $A[i]$, for every $i$.
Then, we have queries supported in $O(1)$, but updates in $O(n)$.
These two solutions represent opposite extreme cases:
the first achieving fastest {\Update} but slowest {\Sum},
the second achieving fastest {\Sum} but slowest {\Update}.
{(They coincide only when $n$ is bounded by a constant.)}
However, it is desirable to have a balance
between the running times of {\Sum} and {\Update}.
One such trade-off
can be obtained by tree-shaped data structures,
whose analysis is the scope of our paper.

An elegant solution is to superimpose on $A$
a balanced binary tree. The leaves of the tree store the elements
of $A$, whereas the internal nodes store the sum of the elements
of $A$ descending from the left and right sub-trees.
As the tree is balanced and has height $\lceil \log_2 n \rceil + 1$,
it follows that both {\Sum} and {\Update} translate into tree
traversals with $O(1)$ spent per level.
This data structure -- called {\segtree}~\cite{bentley1977solutions} --
guarantees $O(\log n)$ for both queries and updates.
Another tree layout having $O(\log n)$-complexity
is the {\fentree}~\cite{fenwick1994new}.
Differently from the {\segtree}, the {\fentree} is an \emph{implicit}
data structure, i.e., it consumes exactly $n+1$ memory words for
representing $A$ is some appropriate manner.
It is not a binary tree and exploits bit-level programming tricks
to traverse the implicit tree structure.

Interestingly, it turns out that a logarithmic complexity is optimal
for the problem
{when $\delta$ is as large as
the machine word~\cite{patrascu2006logarithmic}.}
Thus, it is interesting to design
efficient implementations of both {\segtree} and {\fentree}
to understand what can be achieved in practice.

Furthermore, it is natural to generalize the {\segtree} to become $b$-ary
and have a height of $\lceil \log_b n \rceil$,
with $b>2$:
an internal node stores an array $B$ of $b-1$ values,
where $B[i]$ is the sum of the elements of $A$ covered by
the sub-tree rooted in its $i$-th child.
Now, we are concerned with solving a smaller instance of the original problem,
i.e., the one having $B$ as input array.
According to the solution adopted for the ``small array'',
different complexities can be achieved.
To give an idea, one could just adopt
one of the two trivial solutions discussed above.
If we leave the elements of $B$ as they are, then we
obtain updates in $O(\log_b n)$ and queries in $O(b \log_b n)$.
Conversely, if we pre-compute
the answers to {\Sum}, we have $O(b \log_b n)$ for updates and
$O(\log_b n)$ for queries.

As a matter of fact,
essentially all theoretical constructions
are variations of a $b$-ary {\segtree}~\cite{dietz1989optimal,raman2001succinct,hon2011succinct,patrascu2006logarithmic}.
An efficient implementation of such data structure is, therefore,
not only interesting for this reason but also particularly appealing
in practice because it opens the possibility of using SIMD instructions
to process in parallel the $b$ keys stored at each node of the tree.
Lastly, also the {\fentree} extends to branching factors larger than 2,
but is a less obvious way that we discuss later in the paper.

\paragraph*{Contributions}
For all the reasons discussed above,
we describe efficient implementations of and compare
the following solutions:
the {\segtree}, the {\fentree}, the $b$-ary {\segtree},
and the $b$-ary {\fentree} -- plus other optimized variants that we will
introduce in the rest of the paper.
We show that,
by taking into account
(1) branch-free code execution,
(2) cache-friendly node layouts,
(3) SIMD instructions, and
(4) compile-time optimizations,
new interesting trade-offs can be established on modern hardware.

After a careful experimental analysis, we arrive at the conclusion
that an optimized $b$-ary {\segtree} is the best data structure.
Very importantly, we remark that optimizing the {\segtree}
is not only relevant for the prefix-sum problem, because
this data structure can also be used to solve several other
problems in computational geometry, such as
\emph{range-min/max queries}, \emph{rectangle intersection},
\emph{point location}, and \emph{three-sided queries}.
(See the book by~\citet{BergCKO08}
and references therein for an introduction to such problems.)
Thus, the contents of this paper can be adapted to solve these
problems as well.

To better support our exposition, we directly show (almost) full
C++ implementations of the data structures, in order to
guide the reader into a deep performance tuning of the software.
We made our best to guarantee that the presented code results compact
and easy to understand but without, for any reason,
sacrificing its efficiency.

The whole code used in the article
is freely available at
\url{https://github.com/jermp/psds}, with detailed
instructions on how to run the benchmarks and reproduce the results.


\section{Related Work}\label{sec:related_work}

Literature about the prefix-sum problem is rich of theoretical results
that we summarize here.
These results are valid under a RAM model with word size $w$ bits.
Let us denote with $t_q$ and $t_u$
the worst-case complexities for queries and updates respectively.

The first important result was given by~\citet*{FS89}, who proved a lower bound
of $\Omega(\log n/\log w)$, which is $\Omega(\log n/ \log\log n)$ for the
typical case where an integer fits in one word, i.e., $w=\Theta(\log n)$.
They also gave the trade-off $t_q \log_2(w t_u) = \Omega(\log n)$.
The same lower bound was found by~\citet*{yao1985complexity}
using a different method.
\citet*{hampapuram1998optimal} gave a $\Omega(\log n)$ lower bound
for the amortized complexity.
\citet*{dietz1989optimal} proposed a data structure that achieves \citet*{FS89}'s
lower bound for both operations.
However, his solution requires $\delta$ to be $O(\log\log n)$
(we recall that $\delta$ is the number of bits needed to represent $\Delta$).
He designed a way to handle $b = O(\log^{\varepsilon} n)$ elements in constant time,
for some $\varepsilon > 0$, and then built a {\segtree} with branching factor $b$.
Thus, the height of the tree is $O(\log_b n) = O(\log n/\log\log n)$.
To handle $b$ elements in constant time, the prefix sums are computed into
an array $B$, whereas the most recent updates are stored into another array $C$.
Such array $C$ is handled in constant time using a precomputed table.
This solution has been improved by~\citet{raman2001succinct} and then,
again, by~\citet{hon2011succinct}.
The downside of these solutions is that they require large universal tables
which do not scale well with $\delta$ and $b$.

\citet*{patrascu2006logarithmic} considerably improved the previous lower bounds
and trade-off lower bounds for the problem. In particular, they determined
a parametric lower bound in $w$ and $\delta$.
Therefore, they distinguished two cases for the problem:
the case where $\delta = \Omega(w)$ bits and the case where $\delta = o(w)$ bits.

In the former case they found the usual logarithmic bound
$\max\{t_q,t_u\} = \Omega(\log n)$.
They also found two symmetrical trade-off lower bounds:
$t_q \log_2(t_u/t_q) = \Omega(\log n)$ and $t_u \log_2(t_q/t_u) = \Omega(\log n)$.
This proves the optimality of the {\segtree} and {\fentree}
data structures introduced in Section~\ref{sec:introduction}.

In the latter case they found a trade-off lower bound of
$t_q(\log_2(w/\delta)+\log_2(t_u/t_q)) = \Omega(\log n)$
which implies the lower bound $\max\{t_q,t_u\} = \Omega(\log n/\log(w/\delta))$.
(Note that in the case where $\delta = \Theta(\log n)$
it matches the lower bounds for the first case.)
They also proposed a data structure for this latter case that achieves
$t_q = t_u = O(\log n/\log(w/\delta))$ which
is optimal for the given lower bound.
Their idea is to support both operations in constant time on $b$ 
elements and build a {\segtree} with branching factor $b$, as already done
by Dietz.
Again, the prefix sums for all the elements are precomputed and
stored in an array $B$. The recent updates are kept in another array $C$.
In particular, and differently from Dietz's solution,
the array $C$ is packed in $w$ bits in order to manipulate it in constant time,
which is possible as long as $b = O(w/(\delta+\log w))$.

In Section~\ref{sec:segtree_bary} we will design an efficient
and practical implementation of a $b$-ary {\segtree}, which also takes 
advantage of smaller bit-widths $\delta$ to enhance the runtime.

The lower bounds found by~\citet*{patrascu2006logarithmic} do not prevent
one of the two operations to be implemented in time $o(\log n)$.
Indeed, \citet*{chan2010counting} proposed a data structure that achieves 
$t_q = O(\log n/\log\log n)$ but $t_u = O(\log^{0.5+\varepsilon} n)$,
for some $\varepsilon > 0$.
Their idea is (again) to build a {\segtree} with branching factor $b = \sqrt{w}$.
Each node of the tree is represented as another tree, but with branching factor
$b^{\prime} = \varepsilon\log w$.
This smaller tree supports {\Sum} in $O(b/b^{\prime})$ time and {\Update} in
$O(1 + 2^{b^{\prime}} b^2/w)$ amortized time.
To obtain this amortized bound, each smaller tree keeps an extra machine word
that holds the $k = w/b$ most recent updates.
Every $k$ updates, the values are propagated to the children of the node.


\paragraph*{Some Remarks}
First, it is important to keep in mind that a theoretical solution
may be too complex and, hence, of little practical use,
as the constants hidden in the asymptotic bounds are high.
For example,
we tried to implement the data structure proposed by~\citet*{chan2010counting},
but soon found out that the algorithm
for {\Update} was too complicated and actually performed much worse than
a simple $O(\log n)$-time solution.

Second, some studies tackle the problem of reducing the space
of the data structures~\cite{raman2001succinct,hon2011succinct,bille2017succinct,vigna2019};
in this article we do not take into account
compressed representations\footnote{
As a reference point, we determined that the fastest compressed {\fentree}
layout proposed by~\citet*{vigna2019}, the so-called \textsf{byte[F]} data structure
described in the paper, is not faster than the classic uncompressed {\fentree}
\emph{when} used to support {\Sum} and {\Update}.}.

Third, a different line of research studied the problem
in the parallel-computing setting where the solution is computed
using $p>1$ processors~\cite{meijer1987optimal,goodrich1994optimal}
or using specialized/dedicated
hardware~\cite{harris2007parallel,brodnik20061}.
As already mentioned in Section~\ref{sec:introduction}, we entirely
focus on single-core solutions that should run on commodity hardware.

Lastly, several extensions to the problem exist, such as
the so-called
\emph{searchable prefix-sum problem}~\cite{raman2001succinct,hon2011succinct},
where we are also asked to support the operation $\Search(x)$ which returns
the smallest $i$ such that $\Sum(i) \geq x$;
and the \emph{dynamic prefix-sum problem} with insertions/deletions of elements
in/from $A$ allowed~\cite{BCGSVV15}.


\section{Experimental Setup}\label{sec:setup}

Throughout the paper we show experimental results,
so we describe here our experimental setup and methodology.

\begin{table}[t]
\centering
\caption{Cache hierarchy on the Intel i9-9940X processor.
All cache levels have a line size of 64 bytes.}
\scalebox{1.0
}{\begin{tabular}{c c c r r}
\toprule
Cache & Associativity & Accessibility & KiB & 64-bit Words \\
\midrule


$L_1$ &  8 & private &     32 &     4,096 \\
$L_2$ & 16 & private &  1,024 &   131,072 \\
$L_3$ & 11 & shared  & 19,712 & 2,523,136 \\

\bottomrule
\end{tabular}
}
\label{tab:caches}
\end{table}

\paragraph*{Hardware}
For the experiments reported in the article we used an Intel i9-9940X processor,
clocked at 3.30 GHz.
The processor has two private levels of cache memory per core:
$2 \times 32$ KiB $L_1$ cache
(32 KiB for instructions and 32 KiB for data); 1 MiB for $L_2$ cache.
A $L_3$ cache level spans $\approx$19 MiB and is shared among all cores.
Table~\ref{tab:caches} summarizes these specifications.
The processor supports the following SIMD~\cite{SIMDIntel} instruction sets:
MMX, SSE (including 2, 3, 4.1, 4.2, and SSSE3),
AVX, AVX2, and AVX-512.

{(We also confirmed our results using another Intel i9-9900X processor,
clocked at 3.60 GHz with the same $L_1$ configuration,
but smaller $L_2$ and $L_3$ caches.
Although timings were slightly different, the same conclusions held.)}

\paragraph*{Software}
All solutions described in the paper were implemented in C++.
The code is available at
\url{https://github.com/jermp/psds}
{and was tested on Linux with \texttt{gcc} 7.4 and 9.2,
and Mac OS with \texttt{clang} 10 and 11.}
For the experiments reported in the article, the code was compiled with
\texttt{gcc} 9.2.1 under Ubuntu 19.10 (Linux kernel 5.3.0, 64 bits),
using the flags:
\texttt{-std=c++17 -O3 -march=native -Wall -Wextra}.

\paragraph*{Methodology}
All experiments run on a \emph{single} core of the processor, with the
data residing entirely in memory so that disk operations
are of no concern. (The host machine has 128 GiB of RAM.)
{Performance counts, e.g., number of performed instructions,
cycles spent, and cache misses, were collected for all algorithms,
using the Linux \textsf{perf} utility. We will use such performance counts
in our discussions to further explain the behavior of the algorithms.}

We operate in the most general setting, i.e., with no specific
assumptions on the data:
arrays are made of 64-bit signed integers and initialized at random;
the $\Delta$ value for an {\Update} operation is also a random
signed number whose bit-width is 64 unless otherwise specified.

To benchmark the speed of the operations for a given array size $n$,
we generate $10^4$
(pseudo-) random natural numbers in the interval $[0,n)$ and use these
as values of $i$ for both $\Sum(i)$ and $\Update(i,\Delta)$.
Since the running time of {\Update} does not depend on
the specific value of $\Delta$, we set $\Delta = i$ for the updates.

We show the running times for {\Sum} and {\Update} using plots
obtained by varying $n$ from a few hundreds up to one billion integers.
In the plots the minimum and maximum timings obtained by varying $n$
draw a ``pipe'', with the marked line inside representing
the average time.
The values on $n$ used are rounded powers of $10^{1/10} \approx 1.25893$:
this base was chosen so that data points are evenly spaced
when plotted in a logarithmic scale~\cite{khuong2017array}.
All running times are reported in nanoseconds spent per operation
(either, nanosec/{\Sum} or nanosec/{\Update}).
Prior to measurement, a warm-up run is executed to fetch the
necessary data into the cache.

%

\section{The Segment-Tree}\label{sec:segtree}

The {\segtree} data structure was originally proposed by~\citet*{bentley1977solutions}
in an unpublished manuscript, and later formalized by~\citet*{bentley1980optimal}
to solve the so-called \emph{batched range problem} (given a set of rectangles
in the plane and a query point, report all the rectangles where the point lies in).

\begin{figure}[t]
\centering

\subfloat[]{
\includegraphics[scale=1]{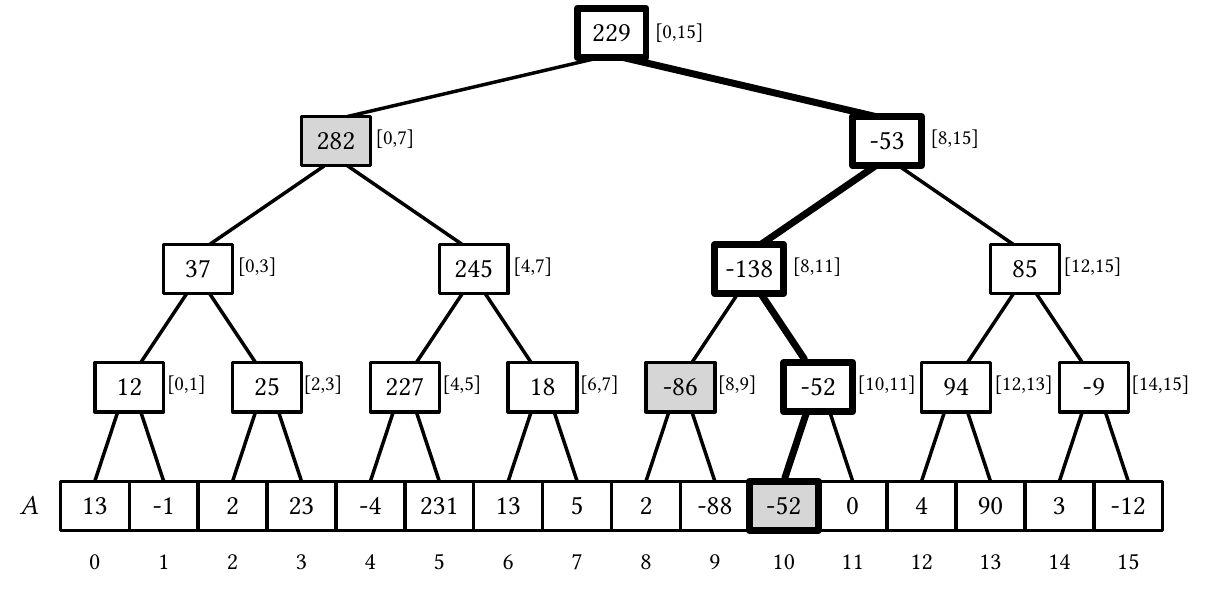}
\label{fig:segtree}
}

\subfloat[]{
\includegraphics[scale=0.95]{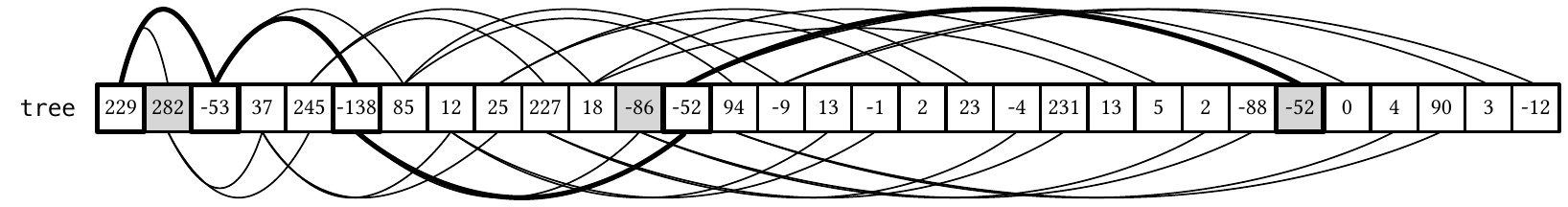}
\label{fig:implicit_tree}
}

\caption{In (a), an example of {\segtree} built from an input array $A$,
with highlighted nodes belonging to the root-to-leaf path for answering $\Sum(10)$.
The shaded nodes are the ones accessed to compute the
result, thus $\Sum(10)=282-86-52=144$.
In (b), the array-like representation of the same tree.
}

\end{figure}

Given an array $A[0..n)$, the {\segtree} is a complete balanced binary tree
whose leaves correspond to the individual elements
of $A$ and the internal nodes hold the sum of the elements of $A$
descending from the left and right sub-trees.
The ``segment'' terminology derives from the fact that each internal node
logically covers a segment of the array and holds the sum of the
elements in the segment.
Therefore, in simplest words, the {\segtree} is a hierarchy
of segments covering the elements of $A$.
In fact, the root of the tree covers the entire array, i.e.,
the segment $[0,n-1]$;
its children cover half array each, that is
the segments $[0,\lfloor (n-1)/2 \rfloor]$ and $[\lfloor (n-1)/2 \rfloor + 1,n-1]$
respectively, and so on until the $i$-th leaf spans the segment $[i,i]$
which corresponds to $A[i]$.
Figure~\ref{fig:segtree} shows a graphical example for an array of
size 16.

\paragraph*{Top-Down Traversal}
Both {\Sum} and {\Update} can be resolved
by traversing the tree structure \emph{top-down} with a classic
binary-search algorithm.
For example, Figure~\ref{fig:segtree} highlights the nodes traversed
during the computation of $\Sum(10)$.
To answer a $\Sum(i)$ query, for every traversed node
we determine the segment comprising $i$
by comparing $i$ with the index in the \emph{middle} of the
segment spanned by the node
and moving either to the left or right child.
Every time we move to the right child, we can sum the value
stored in the left child to our result.
Updates are implemented in a similar way.
Since each child of a node spans half
of the segment of its parent, the tree is balanced and its height
is $\lceil \log_2 n \rceil + 1$.
It follows that {\Sum} and {\Update} are supported in $O(\log n)$ time.

\begin{figure}[t]
\begin{lstlisting}
struct segment_tree_topdown {
    void build(uint64_t const* input, uint64_t n) {
        size = 1ULL << uint64_t(ceil(log2(n)));
        tree.resize(2 * size - 1, 0);
        vector<int64_t> in(size, 0);
        copy(input, input + n, in.begin());
        visit(in.data(), 0, size - 1, 0);
    }

    int64_t sum(uint64_t i) const {
        uint64_t l = 0, h = size - 1, p = 0;
        int64_t sum = 0;
        while (l < h) {
            if (i == h) break;
            uint64_t m = (l + h) / 2;
            p = 2 * p + 1;
            if (i > m) { sum += tree[p]; l = m + 1; p += 1; }
            else h = m;
        }
        return sum + tree[p];
    }

    void update(uint64_t i, int64_t delta) {
        uint64_t l = 0, h = size - 1, p = 0;
        while (l < h) {
            tree[p] += delta;
            uint64_t m = (l + h) / 2;
            p = 2 * p + 1;
            if (i > m) { l = m + 1; p += 1; }            
            else h = m;
        }
        tree[p] += delta;
    }

private:
    uint64_t size; vector<int64_t> tree;

    int64_t visit(int64_t const* in, uint64_t l, uint64_t h, uint64_t p) {
        if (l == h) return tree[p] = in[l];
        uint64_t m = (l + h) / 2;
        int64_t l_sum = visit(in, l, m, 2 * p + 1);
        int64_t r_sum = visit(in, m + 1, h, 2 * p + 2);
        return tree[p] = l_sum + r_sum;
    }
};
\end{lstlisting}
\caption{A top-down implementation of the {\segtree}.
\label{code:segtree}}
\end{figure}


In Figure~\ref{code:segtree} we give the full code implementing
this top-down approach. Some considerations are in order.
First, observe that we build the tree by padding the array
with zeros until we reach the first power of 2 that is
larger-than or equal-to $n$.
This substantially simplifies the code,
and permits further optimizations
that we will introduce later.
Second, we store the entire tree in an array, named \code{tree} in the code,
thus representing the tree topology \emph{implicitly} with
parent-to-child relationships coded via integer arithmetic:
if the parent node is stored in the array at position $i$,
then its left and right children are placed
at positions $2i + 1$ and $2i + 2$ respectively.
Note that this indexing mechanism requires
\emph{every} node of the tree to have two children, and this is (also)
the reason why we pad the array with zeros to reach
the closest power of two.
Figure~\ref{fig:implicit_tree} shows the array-like view of the
tree in Figure~\ref{fig:segtree}.


The complete tree hierarchy, excluding the leaves which correspond
to the elements of the original array,
consists of $2^{\lceil \log_2 n \rceil}-1$ internal nodes.
These internal nodes are stored in the first half of the array,
i.e., in \code{tree[0..size-1)};
the leaves are stored in the second half,
i.e., in \code{tree[size-1..2*size-1)}.
Thus the overall {\segtree} takes a
total of $2^{\lceil \log_2 n \rceil+1}-1$ 64-bit words.
Therefore, we have that the {\segtree} takes $c \cdot n$ memory words,
with $c \in [2,4)$. The constant $c$ is 2 when $n$ is a power of 2,
but becomes close to 4 when $n$ is very distant from
$2^{\lceil \log_2 n \rceil}$.

\paragraph*{Bottom-Up Traversal}
In Figure~\ref{code:segtree_bottomup} we show an alternative
implementation of the {\segtree}.
Albeit logically equivalent to the code
shown in Figure~\ref{code:segtree},
this implementation has advantages.

First, it avoids extra space. In particular,
it always consumes $2n-1$ memory words,
\emph{while preserving implicit parent-to-child
relationships} expressed via integer arithmetic.
Achieving this when $n$ is a power of 2 is trivial because
the tree will always be full, but is more difficult otherwise.
In the \code{build} method of the class
we show an algorithm that does so.
The idea is to let the leaves of the tree, which
correspond to the elements of the input, be stored in the array \code{tree}
slightly out of order.
The leaves are still stored in the second half of the array but
instead of being laid out consecutively from position
\code{n-1} (as it would be if \code{n} were a power of 2),
they start from position \code{begin}, which can be larger than \code{n-1}.
Let \var{m} be $2n-1$.
The first \var{m - begin} leaves are stored in
\code{tree[begin..m)}; all remaining leaves are stored in \code{tree[n-1,begin)}.
This is achieved with the two \code{for} loops in the
\code{build} method.
(Note that \code{begin} is always $2^{\lceil \log_2 n \rceil}-1$ which is
$n-1$ if $n$ is a power of 2.)
Such \emph{circular} displacement of the leaves guarantees that
parent-to-child relationships are preserved even when $n$
is \emph{not} a power of 2.
Lastly, the \code{visit} method traverses the internal nodes recursively,
writing in each node the proper sum value.

\begin{figure}[t]
\begin{lstlisting}
struct segment_tree_bottomup {
    void build(int64_t const* input, uint64_t n) {
        uint64_t m = 2 * n - 1;
        tree.resize(m);
	size = n;
        begin = (1ULL << uint64_t(ceil(log2(n)))) - 1;
        uint64_t i = 0, j = 0;
        for (; begin + i != m; ++i) tree[begin + i] = input[i];
        for (; i != n; ++i, ++j)       tree[n - 1 + j] = input[i];
        visit(0);
    }

    int64_t sum(uint64_t i) const {
        uint64_t p = leaf(i);
        int64_t sum = tree[p];
        while (p) {
            if ((p & 1) == 0) sum += tree[p - 1]; // p is right child
            p = (p - 1) / 2;
        }
        return sum;
    }

    void update(uint64_t i, int64_t delta) {
        uint64_t p = leaf(i);
        while (p) {
            tree[p] += delta;
            p = (p - 1) / 2;
        }
    }

private:
    uint64_t size, begin; vector<int64_t> tree;

    uint64_t leaf(uint64_t i) const {
        uint64_t p = begin + i;
        p -= (p >= tree.size()) * size;
        return p;
    }

    int64_t visit(uint64_t p) {
        if (p >= m_tree.size()) return 0;
        if (p >= m_size - 1) return m_tree[p];
        int64_t l_sum = visit(2 * p + 1);
        int64_t r_sum = visit(2 * p + 2);
        return m_tree[p] = l_sum + r_sum;
    }
};
\end{lstlisting}
\caption{A bottom-up implementation of the {\segtree}.
\label{code:segtree_bottomup}}
\end{figure}

Second, we now traverse
the tree structure \emph{bottom-up}, instead of top-down.
This direction
allows a much simpler implementation of the {\Sum}
and {\Update} procedures. The inner \code{while} loop is shorter
and it is only governed by the index \var{p}, which is
initialized to be the position of the \var{i}-th leaf using
the function \code{leaf}, and updated
to become the index of the parent node at each iteration
until it becomes \var{0},
the index of the root node.
Furthermore in the code for {\Sum},
every time \var{p} is the index of a \emph{right} child
we sum the content of its left sibling, which is stored in
\var{tree[p-1]}, to the result.

To check whether \var{p} is the index of a right child,
we exploit the property that
\emph{left children are always stored at odd positions in the array;
right children at even positions}.
This can be proved by induction on \var{p}.
If \var{p} is 0, i.e., is the index of the root, then
its left child is in position 1 and its right child in position 2,
hence the property is satisfied.
Now, suppose it holds true when $\var{p} = k$, for some $k > 0$.
To show that the property holds for the children of \var{p}, it is
sufficient to recall that: the double of a number is always even;
if we sum 1 to an even number, the result is an odd number (left child);
if we sum 2 to an even number, the result is an even number (right child).
In conclusion, if the parity of \var{p} is 0, it must be a right child.
(The parity of \var{p} is indicated by its first bit from the right,
that we isolate with \code{p \& 1}.)

%

\paragraph*{Branch-Free Traversal}
Whatever implementation of the {\segtree} we use,
either top-down or bottom-up,
a branch (\code{if} statement)
is always executed in the \code{while} loop of
{\Sum} (and in that of {\Update} for the top-down variant).
For randomly-distributed queries, we expect the branch
to be hard to predict: it will be true for approximately half
of the times, and false for the others.
Using speculative execution as branch-handling policy,
the penalty incurred whenever the processor mispredicts a branch
is a \emph{pipeline flush}: all the instructions executed so far
(speculatively) must be discarded from the processor pipeline,
for a decreased
instruction throughput. Therefore, we would like to avoid
the branch inside the \code{while} loop.


\begin{figure}[t]
\subfloat[top-down]{
\begin{minipage}[t]{0.5\textwidth}
\begin{lstlisting}
int64_t sum(uint64_t i) const {
    uint64_t n = size;
    uint64_t m = (size - 1) / 2, p = 0;
    int64_t sum = 0;
    while (n != 1) {
        uint64_t cmp = i > m;
        p = 2 * p + 1;
        sum += cmp * tree[p];
        p += cmp;
	n /= 2;
        int64_t offset = cmp * n - n / 2;
        m += offset;
    }
    return sum + tree[p];
}
\end{lstlisting}
\label{code:branchfree-sum-a}
\end{minipage}
}
\subfloat[bottom-up]{
\begin{minipage}[t]{0.5\textwidth}
\begin{lstlisting}
int64_t sum(uint64_t i) const {
    uint64_t p = leaf(i);
    int64_t sum = tree[p];
    while (p) {
        uint64_t cmp = (p & 1) == 0;
        sum += cmp * tree[p - 1];
        p = (p - 1) / 2;
    }
    return sum;
}
\end{lstlisting}
\end{minipage}
}
\caption{Branch-free {\Sum} implementations on the {\segtree}.
\label{code:branchfree-sum}}
\end{figure}

\begin{figure}[t]
\centering

\subfloat[{\Sum}]{
\includegraphics[scale=\myfigsize]{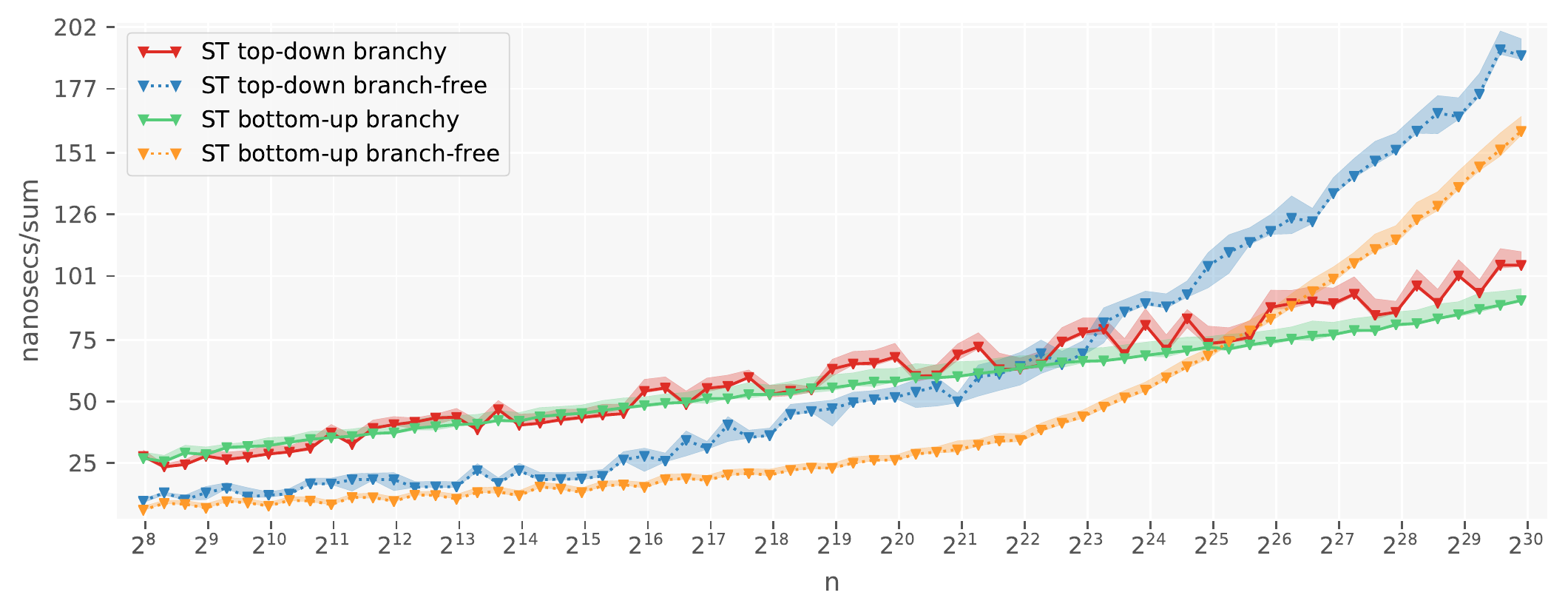}
}

\subfloat[{\Update}]{
\includegraphics[scale=\myfigsize]{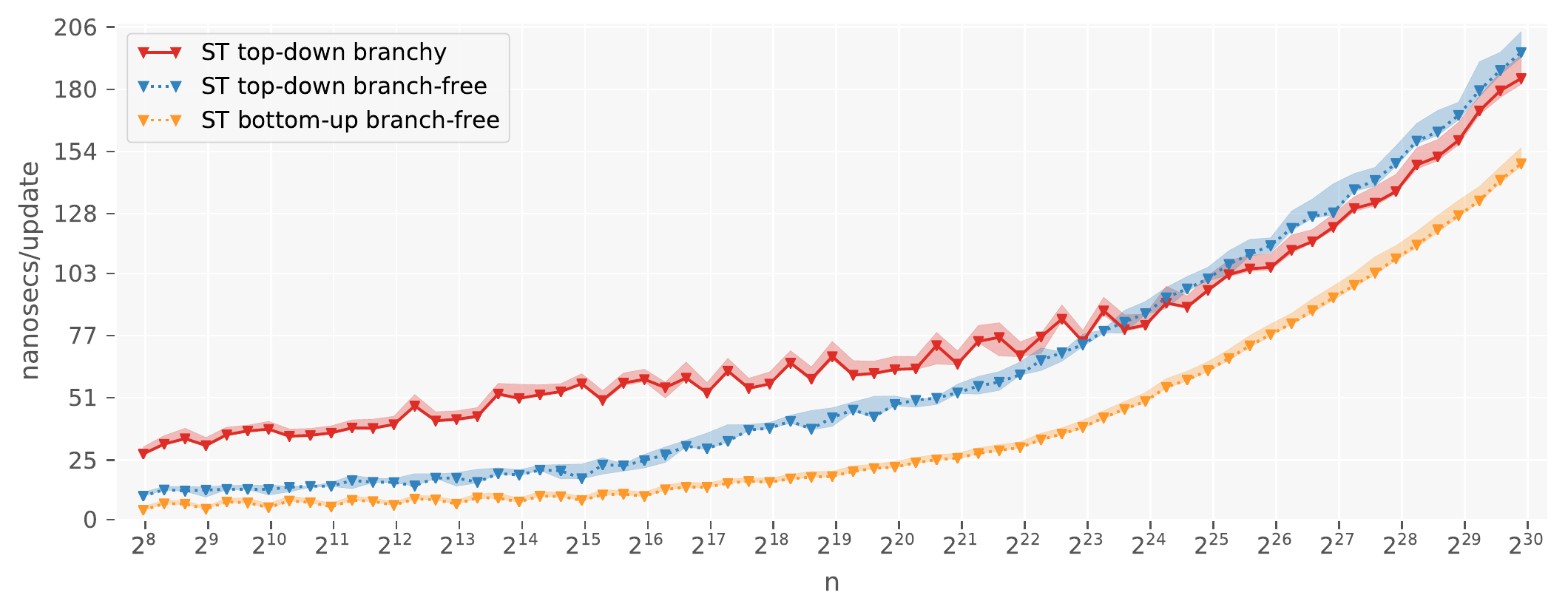}
\label{fig:branchy_vs_branchfree:update}
}

\caption{Running times for branchy and branch-free {\Sum}/{\Update} on the {\segtree} (ST).
\label{fig:branchy_vs_branchfree}}
\end{figure}

In Figure~\ref{code:branchfree-sum} we show a branch-free
implementation of the {\Sum} algorithm\footnote{The {\Update} algorithm
for top-down is completely symmetric to that of {\Sum}, and not shown here.
Also, note that the {\Update} algorithm for the bottom-up variant
is already branch-free.}.
It basically uses the result of the comparison, \var{cmp},
which will be either 1 or 0,
to appropriately mask\footnote{It is interesting to report that, although the code
shown here uses multiplication, the assembly output by the compiler does
not involve multiplication at all.
The compiler implements the operation \code{cmp * tree[p]} with
a \emph{conditional move} instruction (\code{cmov}), which loads
into a register the content of \code{tree[p]} only if \code{cmp} is true.
} the quantity we are summing to the result
(and move to the proper child
obliviously in the top-down traversal).
The correctness is immediate and left to the reader.
The result of the comparison between the two approaches,
branchy vs. branch-free, is shown in Figure~\ref{fig:branchy_vs_branchfree}.

As we can see, the branch-free implementations
of both {\Sum} and {\Update} are much faster
than the branchy counterparts,
on average by $2\times$ or more,
for a wide range of practical values of $n$.
We collected
some performance counts using the Linux \texttt{perf}
utility to further explain this evidence.
Here we report an example for the bottom-up approach and {\Sum} queries.
For $n \approx 2^{16}$,
the branch-free code spends 66\% less cycles than the
branchy code although it actually executes 38\% more instructions,
thus rising the instruction throughput from 0.54 to 2.2.
Furthermore, it executes 44\% less branches
and misses only 0.2\% of these.
Also, note that the bottom-up version is considerably faster
than the top-down version
(and does not sacrifice anything for the branchy implementation).
Therefore, from now on we focus on the bottom-up version of the
{\segtree}.

%
%
%

However, the branchy implementation of {\Sum} is faster
than the branch-free one for larger values of $n$
(e.g., for $n>2^{25}$).
This shift in the trend is due to the fact
that the latency of a memory access dominates the cost of a
pipeline flush for large $n$.
In fact, when the data structures are sufficiently large,
accesses to larger but \emph{slower} memory levels are involved:
in this case,
one can obtain faster running times by avoiding such accesses
when unnecessary.
The branchy implementation does so. In particular,
it performs a memory access \emph{only if} the condition
of the branch is satisfied, thus saving roughly 50\% of the
accesses for a random query workload
compared to the branch-free implementation.
This is also evident for {\Update}.
All the different implementations of {\Update}
perform an access at every iteration of their loop,
and this is the reason why all the curves
in Figure~\ref{fig:branchy_vs_branchfree:update} have
a similar shape.

\paragraph*{Two-Loop Traversal}
In the light of the above considerations,
we would actually like to combine the best of the ``two worlds'' by:
executing branch-free code as long as the cost of
a pipeline flush exceeds the cost of a memory access,
which is the case when the accessed data resides in the
smaller cache levels;
executing branchy code otherwise, i.e., when the memory
latency is the major cost in the running time that happens
when the accesses are directed to slower memory levels.
Now, the way the {\segtree} is linearized in the \code{tree}
array, i.e., the first $n-1$ positions of the array
store the internal nodes in level order and
the remaining $n$ store the leaves of the tree,
is particularly suitable to achieve this.
A node at depth $d$ has a $(1/2)^d$ probability
of being accessed,
for randomly distributed queries (root is at depth $d=0$).
Therefore when we repeatedly traverse the tree,
the first $L_1$ positions of the array that correspond
to the top-most $\lceil \log_2 L_1 \rceil$ levels of the tree
are kept in cache $L_1$; the following $L_2 - L_1$ positions
are kept in $L_2$, and so on,
where $L_k$ indicate the size of the $k$-th cache in data items
(the ``64-bit Words'' column in Table~\ref{tab:caches}
at page~\pageref{tab:caches}).
If \code{T} is the size of the prefix of the array \code{tree}
that fits in cache, it is intuitive that,
as long as $\code{p} > \code{T}$, we should prefer the
branchy code and save memory accesses;
vice versa,
when $\code{p} \leq \code{T}$, memory accesses are relatively cheap
and we should opt for the branch-free code.
The following code shows this approach.
\begin{lstlisting}
int64_t sum(uint64_t i) const {
    uint64_t p = leaf(i);
    int64_t sum = tree[p];
    while (p > T) {
        if ((p & 1) == 0) sum += tree[p - 1];  // branchy
        p = (p - 1) / 2;
    }
    while (p) {
        sum += ((p & 1) == 0) * tree[p - 1];  // branch-free
        p = (p - 1) / 2;
    }
    return sum;
}
\end{lstlisting}

The value of the threshold \code{T} governs the number of
loop iterations, out of $\lceil \log_2 n \rceil + 1$,
that are executed in a branchy or branch-free manner.
Its value intuitively
depends on the size of the caches and the value of $n$.
For smaller $n$ we would like to set \code{T} reasonably
high to benefit from the branch-free code;
vice versa for larger $n$,
we would like to perform more branchy iterations.
As a rule-of-thumb, we determined that a good choice of
\code{T} is $L_2 - (n > L_3) \times (L_2 - L_1)$,
which is equal to $L_1$ or $L_2$ if $n>L_3$ or not respectively.

It remains to explain how we can apply the ``two-loop'' optimization
we have just introduced to the {\Update} algorithm, since
its current implementation \emph{always} executes an access
per iteration.
To reduce the number of accesses to memory,
we modify the content of the internal nodes
of the tree. If each node stores the sum of the leaves descending from
its \emph{left} subtree only (rather than the sum of
the elements covered by the left \emph{and} right segments),
then a memory access is saved every time
we do not have to update any node in the left subtree.
We refer to this modification of the internal nodes
as a \emph{left-sum} tree hereafter and show the corresponding
implementation in Figure~\ref{code:segment_tree_bottomup_leftsum}.

\begin{figure}[t]
\centering

\subfloat[{\Sum}]{
\includegraphics[scale=\myfigsize]{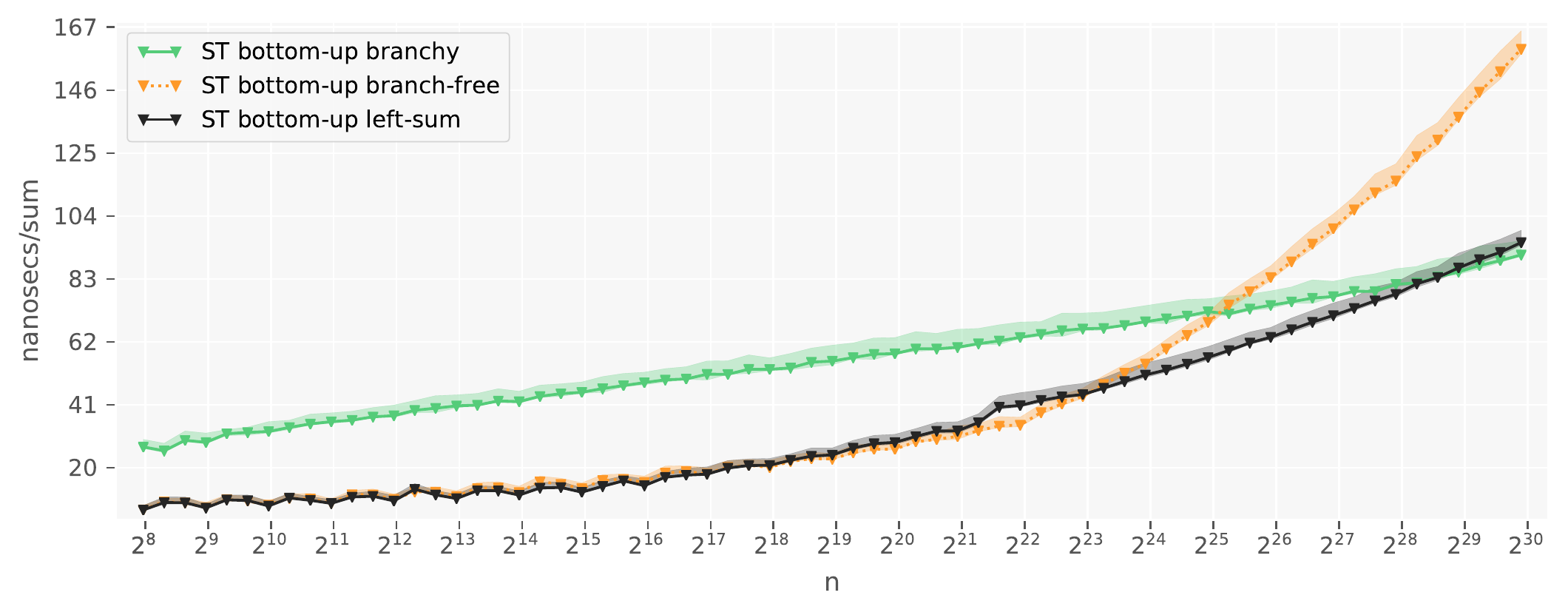}
}

\subfloat[{\Update}]{
\includegraphics[scale=\myfigsize]{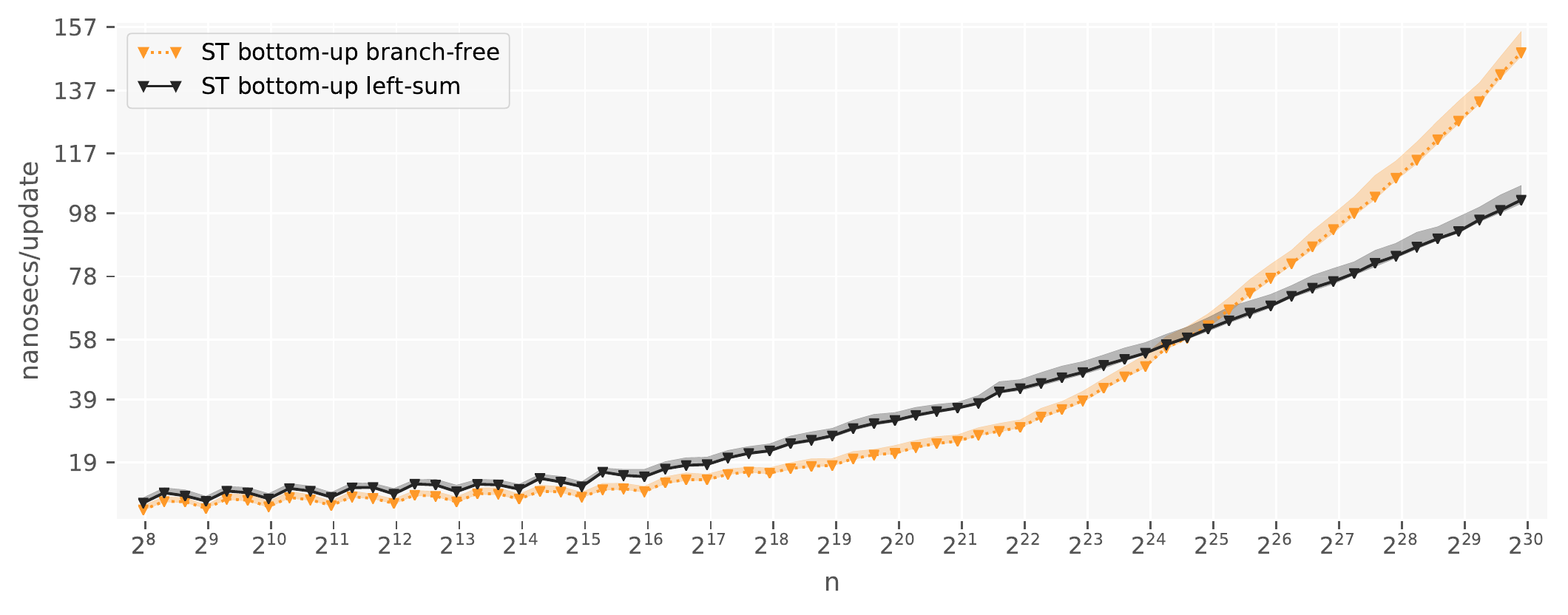}
}

\caption{Comparison between branchy/branch-free implementations
of regular and \emph{left-sum} {\segtree}.
\label{fig:leftsum_switch_on_level}}
\end{figure}

Lastly, Figure~\ref{fig:leftsum_switch_on_level} illustrates
the comparison between branchy/branch-free implementations
of regular and \emph{left-sum} {\segtree}.
As apparent from the plots, the left-sum variant with the
two-loop optimization
is the fastest {\segtree} data structure because it combines
the benefits of branch-free and branchy implementations.
From now on and further comparisons, we simply refer
to this strategy as
\textsf{ST} in the plots.

%
%
%

\begin{figure}[t]
\begin{lstlisting}
struct segment_tree_bottomup_leftsum {
    void build(int64_t const* input, uint64_t n);  // same as in Figure 3

    int64_t sum(uint64_t i) const {
        uint64_t p = leaf(i);
        uint64_t T = L2 - (size > L3) * (L2 - L1);
        int64_t sum = tree[p];
        while (p > T) {
            uint64_t parent = (p - 1) / 2;
            if ((p & 1) == 0) sum += tree[parent];
            p = parent;
        }
        while (p) {
            uint64_t parent = (p - 1) / 2;
            sum += ((p & 1) == 0) * tree[parent];
            p = parent;
        }
        return sum;
    }

    void update(uint64_t i, int64_t delta) {
        uint64_t p = leaf(i);
        uint64_t T = L2 - (size > L3) * (L2 - L1);
        tree[p] += delta;
        while (p > T) {
            uint64_t parent = (p - 1) / 2;
            if ((p & 1) == 1) tree[parent] += delta;
            p = parent;
        }
        while (p) {
            uint64_t parent = (p - 1) / 2;
            tree[parent] += ((p & 1) == 1) * delta;
            p = parent;
        }
    }

private:
    uint64_t size, begin;
    vector<int64_t> tree;

    uint64_t leaf(uint64_t i) const;  // same as in Figure 3

    int64_t visit(uint64_t p) {
        uint64_t l = 2 * p + 1;
        if (l >= tree.size()) return tree[p];
        int64_t l_sum = visit(l), r_sum = visit(l + 1);
        return (tree[p] = l_sum) + r_sum;
    }
};
\end{lstlisting}
\caption{The bottom-up \emph{left-sum} {\segtree} implementation
that stores in each
internal node the sum of the leaves descending from its left sub-tree.
\label{code:segment_tree_bottomup_leftsum}}
\end{figure}

\section{The Fenwick-Tree}\label{sec:fentree}

The solution we describe here is popularly known
as the {\fentree} (or \emph{Binary Indexed Tree}, BIT) after a
paper by~\citet*{fenwick1994new}, although the underlying
principle was introduced by~\citet*{ryabko1992fast}.

The key idea is to exploit the fact that, as a natural number $i > 0$
can be expressed as sum of proper powers of 2, so {\Sum}/{\Update} operations
can be implemented by considering array locations that are power-of-2
elements apart.
For example, since $11 = 2^3+2^1+2^0=8+2+1$, then
$\Sum(10)$ can be computed as $\var{tree}[2^3]+\var{tree}[2^3+2^1]+\var{tree}[2^3+2^1+2^0] = \var{tree}[8]+\var{tree}[10]+\var{tree}[11]$,
where the array \var{tree} is computed from $A$ in some appropriate manner
that we are going to illustrate soon.
In this example, $\var{tree}[8]=\sum_{k=0}^{8-1} A[k]$, $\var{tree}[10]=A[8]+A[9]$, and $\var{tree}[11]=A[10]$.

Now, let $\code{base}=0$. The array $\code{tree}[0..n+1)$
is obtained from $A$ with the following two-step algorithm.
\begin{enumerate}
\item Store in $\var{tree}[\code{base}+i]$ the quantity
$\sum_{k=0}^{i-1} A[\code{base}+k]$
where $i = 2^j$, $j=0,1,2,\ldots$
\item For every sub-array $A[2^j..2^{j+1}-1)$,
repeat step (1) with updated $\code{base}=2^j$.
\end{enumerate}

\begin{figure}[t]
\centering
\includegraphics[scale=1]{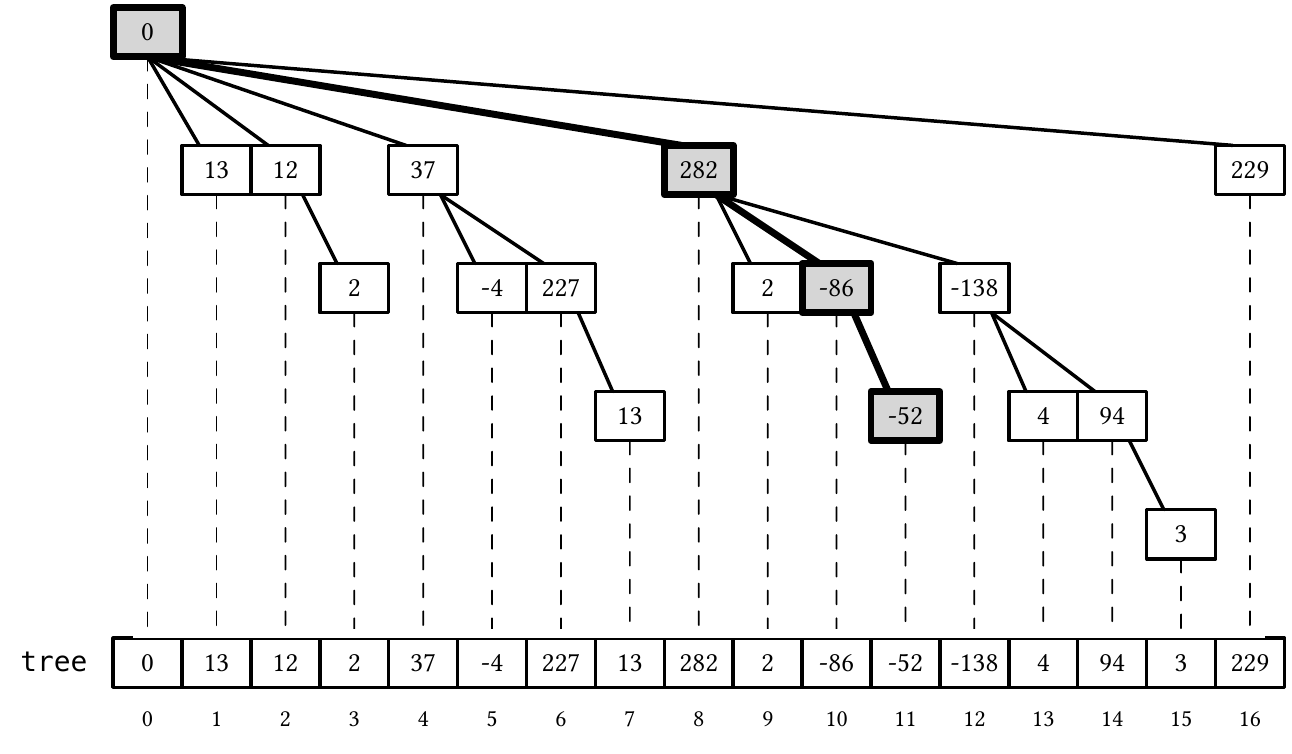}
\caption{An example of the array-like view \var{tree} of the {\fentree},
along with its logical ``interrogation'' tree.
The array \var{tree} is built from the input array
$[$13, -1, 2, 23, -4, 231, 13, 5, 2, -88, -52, 0, 4, 90, 3, -12$]$.
As in Figure~\ref{fig:segtree}, highlighted nodes
belong to the root-to-leaf path that is traversed to answer {\Sum}
for index 10.
\label{fig:interrogation}}
\end{figure}

Figure~\ref{fig:interrogation} shows a tree-like view of one such array \var{tree},
built from the same array $A$ of size 16 shown in Figure~\ref{fig:segtree}.
Note that position 0 is never used for ease of notation
and $\var{tree}[0]=0$, thus \var{tree}'s actual size is $n+1$.
Now, we said that during a {\Sum} query for index $i$ (or {\Update})
we only touch \var{tree}'s positions that are power-of-2 elements apart.
It is not difficult to see that we have one such position for every bit set
in the binary representation of $i$. Since $i$ goes from 0 to $n-1$,
it needs $\lceil \log_2(n+1) \rceil$ bits to be represented, so this will also
be the height of the {\fentree}.
(The depth of a node whose index is $i$ is the Hamming weight
of $i$, i.e., the number of bits set in the binary representation of $i$.)
Therefore, both queries and updates have a worst-case complexity of $O(\log n)$
and the array \var{tree} takes $n+1$ memory words.
In the following we illustrate how to navigate the implicit tree structure
in order to support {\Sum} and {\Update}.

Let us consider the {\Sum} query. As intuitive from the given examples,
when we have to answer $\Sum(i)$, we actually probe
the array \var{tree} starting from index $\code{p}=i+1$.
This is because index 0 is not a power
of 2, thus is can be considered as a dummy entry in the array \var{tree}.
Now, let \code{p} be 11 as in the example of Figure~\ref{fig:interrogation}.
The sequence of nodes' indexes touched during the traversal to compute the result
are (bottom-up): $11 \rightarrow 10 \rightarrow 8 \rightarrow 0$.
It is easy to see that we \emph{always} start from index $i+1$
and end with index 0 (the root of the tree).
To navigate the tree bottom-up we need an efficient way of computing the
index of the parent node from a given index \code{p}.
This operation can be accomplished by
\emph{clearing the least significant bit} (LSB) of
the binary representation of \code{p}.
For example, 11 is \bit{0101\underline{1}} in binary. We underline its LSB.
If we clear the LSB, we get the bit configuration
\bit{010\underline{1}0} which indeed
corresponds to index 10.
Clearing the LSB of 10 gives \bit{0\underline{1}000} which is 8.
Finally, clearing the LSB of a power of 2 will always give 0, that is
the index of the root.
This operation, clearing the LSB of a given number \code{p},
can be implemented efficiently with \code{p \& (p - 1)}.
Again, note that $\Sum(i)$ traverses a number of nodes
equal to the number of bits set (plus 1) in $\code{p}=i+1$.
The code given in Figure~\ref{code:fentree} illustrates the approach.

\begin{figure}[t]
\centering
\includegraphics[scale=1]{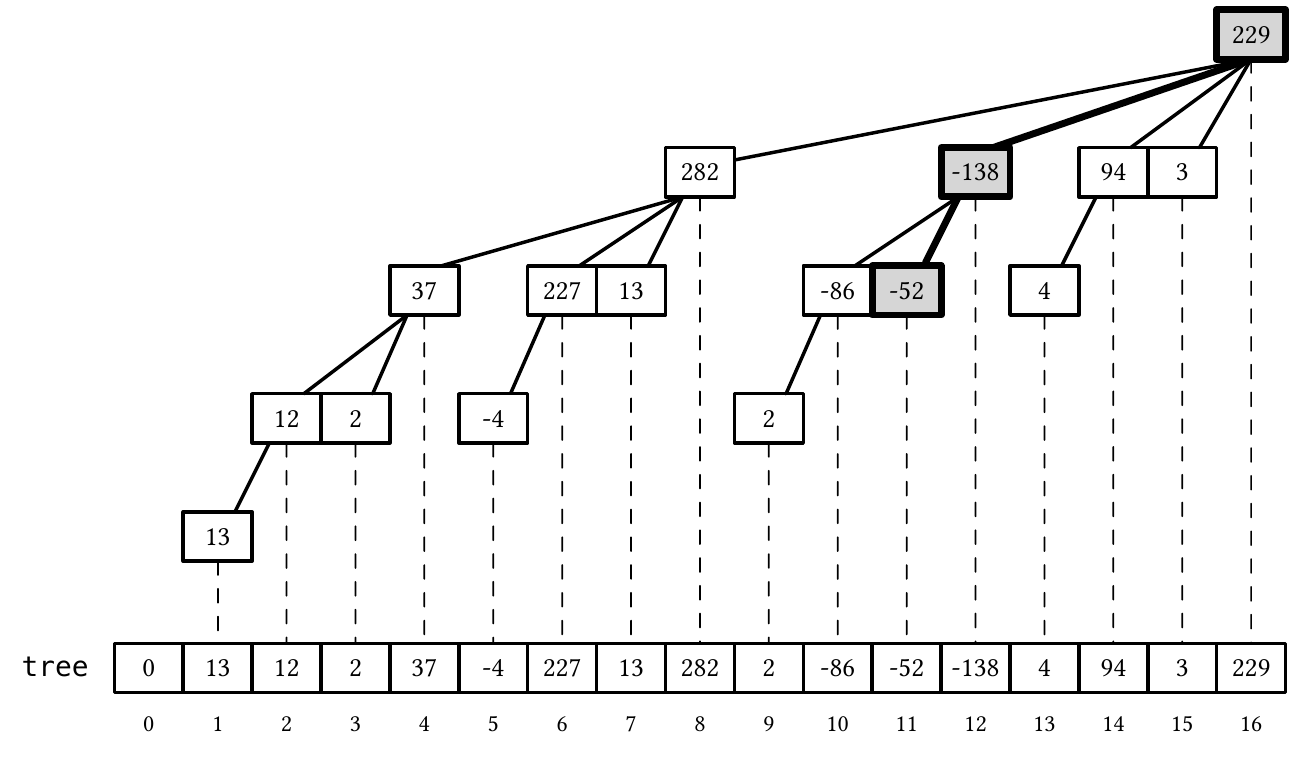}
\caption{The logical ``updating'' tree for the same example
shown in Figure~\ref{fig:interrogation}.
The highlighted nodes
belong to the root-to-leaf path that is traversed to perform {\Update}
for index 10.
\label{fig:updating}}
\end{figure}

The {\fentree} can be actually viewed as the
superimposition of two different trees.
One tree is called the ``interrogation''
tree because it is used during {\Sum}, and it is shown in Figure~\ref{fig:interrogation}.
The other tree is called the ``updating'' tree, and it is shown
in Figure~\ref{fig:updating} instead.
This tree consists of the very same nodes
as the ``interrogation'' tree but
with different child-to-parent relationships.
In fact, starting from
an index $\code{p} = i + 1$ and traversing the tree bottom-up we obtain the
sequence of nodes that need to be updated when issuing $\Update(i,\Delta)$.
Again for the example $i=10$, such sequence is
$11 \rightarrow 12 \rightarrow 16$.
Starting from a given index \code{p}, this sequence can be obtained by
isolating the LSB of \code{p} and summing this to \code{p} itself.
The LSB can be isolated with the operation \code{p \& -p}.
For $\Update(i,\Delta)$, the number of traversed node
is equal to the number of leading zeros (plus 1) in the binary
representation of $\code{p}=i+1$.
The actual code for {\Update} is given in Figure~\ref{code:fentree}.

\begin{figure}[t]
\begin{lstlisting}
struct fenwick_tree {
    void build(uint64_t const* input, uint64_t n) {
        tree.resize(n + 1, 0);
        copy(input, input + n, tree.begin() + 1);
        for (uint64_t step = 2; step <= n; step *= 2)
            for (uint64_t i = step; i <= n; i += step) tree[i] += tree[i - step / 2];
    }

    int64_t sum(uint64_t i) const {
        int64_t sum = 0;
        uint64_t p = i + 1;
        while (p) { sum += tree[p]; p &= p - 1; }
        return sum;
    }

    void update(uint64_t i, int64_t delta) {
        uint64_t p = i + 1;
        while (p < tree.size()) { tree[p] += delta; p += p & -p; }
    }

private:
    vector<int64_t> tree;
};
\end{lstlisting}
\caption{The {\fentree} code.
\label{code:fentree}}
\end{figure}

\begin{figure}[t]
\centering
\includegraphics[scale=0.8]{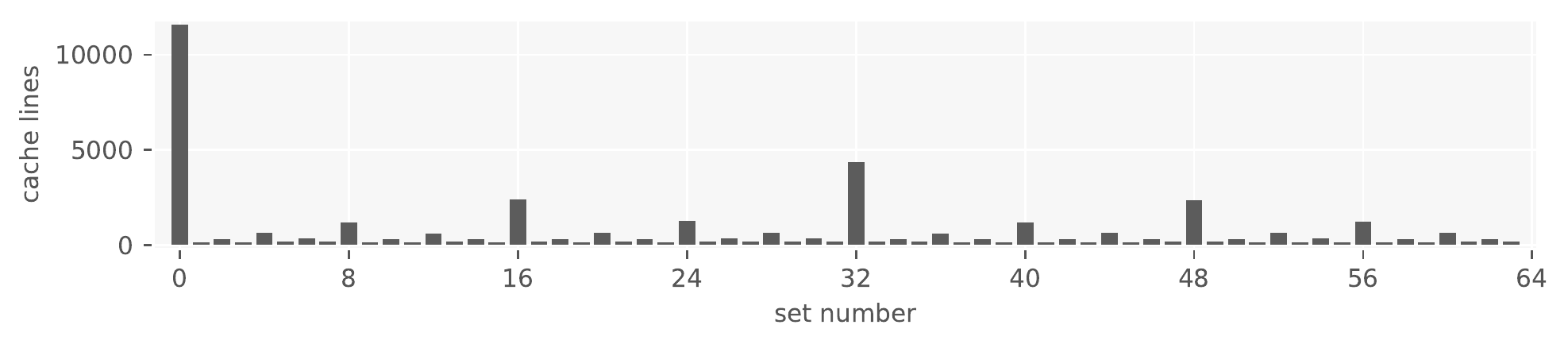}
\caption{Number of distinct cache lines stored in each set
of a $L_1$ 8-way set-associative cache, after running $10^4$ random
{\Sum} queries with a {\fentree}
of size $n=10^7$.
\label{fig:cache_usage_ft}}
\end{figure}

\paragraph*{Cache Conflicts}
The indexing of the nodes in the {\fentree} --
which, for every subtree, places the nodes on the same level
at array locations that are power-of-2 elements apart --
induces a bad exploitation of the processor cache for large
values of $n$.
The problem comes from the fact that cache memories are typically
$c$-way set-associative caches. The cache memory of the processor
used for the experiments in this article is no exception.
In such cache architecture,
a cache line must be stored \emph{in one} (and only one)
set and, if many different cache lines must be stored in the same set,
the set has only room for $c$ of these.
In fact, when the set fills up, a cache line must be evicted from the set.
If the evicted line is then accessed again during the computation
a cache miss will be generated because the line is not in the cache
anymore.
Therefore, accessing (more than $c$) different memory lines
that must be stored
in the same cache set will induce cache misses.
To understand why this happens in the {\fentree}, let us
consider how the set number is determined from a memory address $a$.

Let $C$ be the total cache size in bytes, where each line spans 64 bytes.
The cache stores its lines as divided into $C/(c \times 64)$ sets,
where each set can store a cache line in $c$ different possible
``ways''.
For example, the $L_1$ cache of the processor used for the experiments
in this article
(see Table~\ref{tab:caches} at page~\pageref{tab:caches})
has 8 ways and a total of $C=\num{32768}$ bytes. Therefore there are
$\num{32768} / (8 \times 64) = 64$ sets in the cache, for a total of
512 cache lines.
The first 6 bits (0-5) of $a$ determine the offset into the cache line;
the following 6 bits (6-11) specify the set number.
Thus, the first line of a memory block is stored in the set 0,
the second line is stored in set 1, ecc.
The 64-th line will be stored again in set 0, the 65-th line in set 1, ecc.
It is now easy to see that accessing memory addresses
that are multiple of $64 \times 64 = \num{4096}$ bytes (a memory \emph{page})
is not cache efficient,
since the lines will contend the very same set.

Therefore, accessing memory locations whose distance in memory is a large
power of 2 (multiple of $2^{12}=4096$)
is not cache-friendly.
Unfortunately, this is what happens in the {\fentree} when $n$ is large.
For example, all the nodes at the first level are stored at indexes that
are powers of 2. Thus they will all map to the set 0.
In Figure~\ref{fig:cache_usage_ft} we show the number of distinct
cache lines that must be stored in each set of the cache $L_1$
(8-way set-associative with 64 sets), for $10^4$ random {\Sum}
queries and $n=10^7$. For a total of $\approx$4$\times 10^4$
cache lines accessed, 29\% of these are stored in set 0.
This highly skewed distribution is the source of cache inefficiency.
(Updates exhibit the same distribution.)
Instead, if all accesses were \emph{evenly} distributed among all sets,
we would expect each set to contain $\approx$625 lines (1.56\%).

\begin{figure}[t]
\centering

\subfloat[modified {\fentree}]{
\includegraphics[scale=0.8]{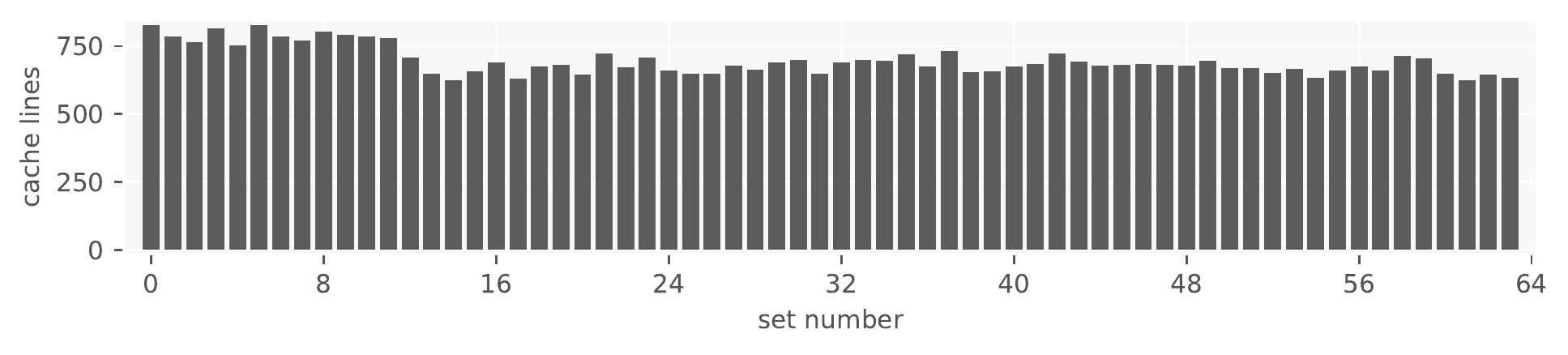}
}

\subfloat[{\segtree}]{
\includegraphics[scale=0.8]{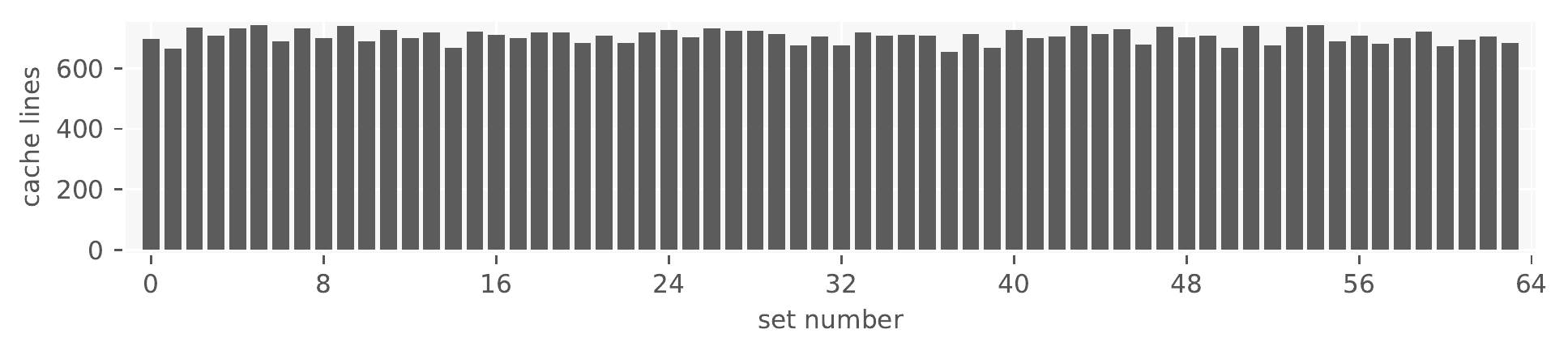}
}

\caption{The same experiment as performed in Figure~\ref{fig:cache_usage_ft},
but after
reducing the cache conflicts in the {\fentree} (modified).
For comparison, we also report the cache usage of the {\segtree}.
\label{fig:cache_usage}}
\end{figure}

This problem can be solved by inserting some \emph{holes} in the
\code{tree} array~\cite{vigna2019},
one every $d$ positions, to let a node whose
position is $i$ in the original array to be placed
at position $i+\lfloor i/d \rfloor$.
This only requires the \var{tree} array to be enlarged
by $\lfloor n/d \rfloor$ words and to recalculate the index of every node
accordingly. If $d$ is chosen to be a sufficiently large
constant, e.g., $2^{14}$, then the extra space is very small.
In Figure~\ref{fig:cache_usage} we show the result of the same
experiment as in Figure~\ref{fig:cache_usage_ft}, but after
the modification to the \var{tree} array and
the new indexing of the nodes.
As it is evident, now every cache set is equally used.
(For comparison, we also report the behavior of the {\segtree}
to confirm that also in this case there is a good usage
of the cache.)

\begin{figure}[t]
\centering

\subfloat[{\Sum}]{
\includegraphics[scale=\myfigsize]{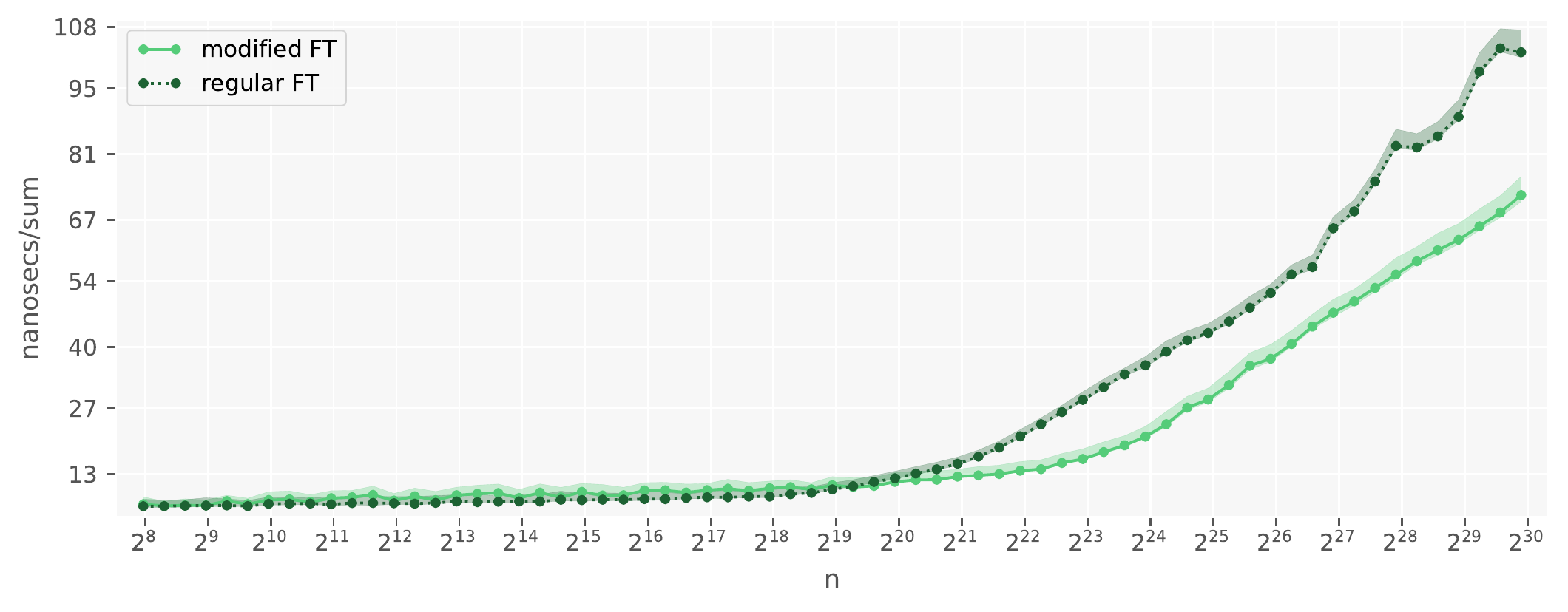}
}

\subfloat[{\Update}]{
\includegraphics[scale=\myfigsize]{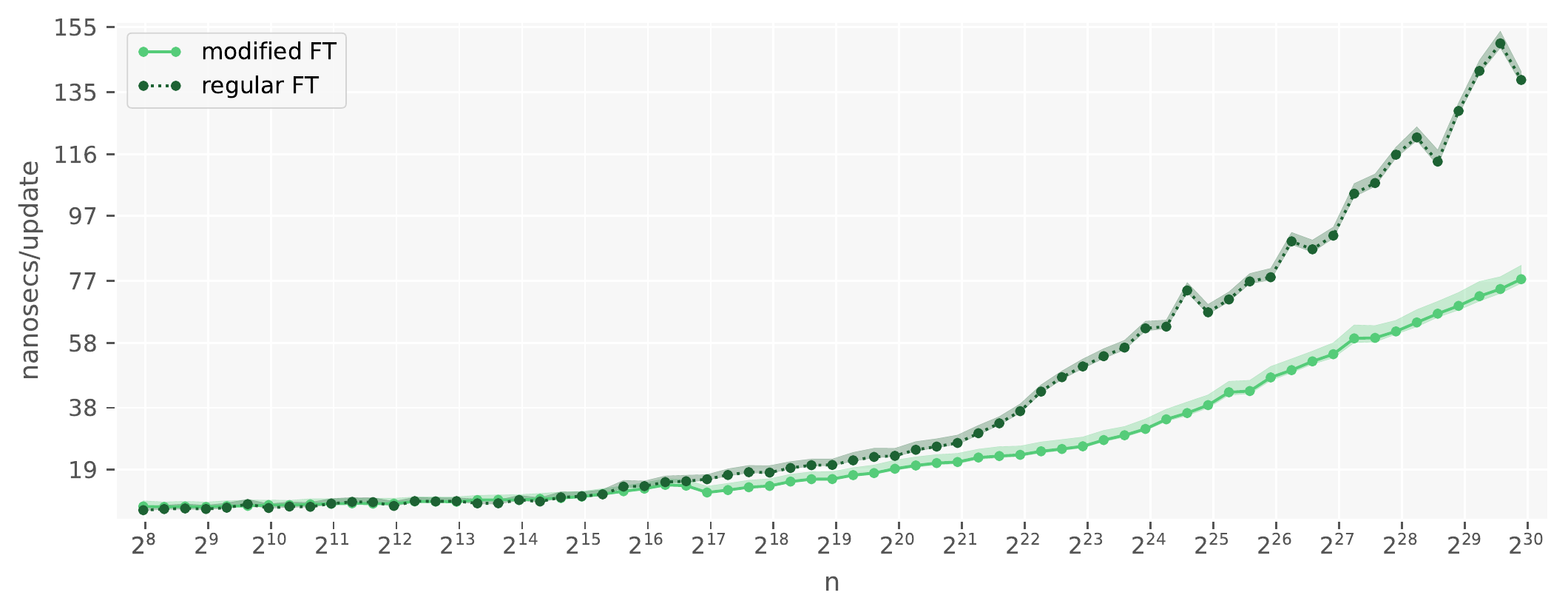}
}

\caption{The running times of {\Sum}/{\Update}
for the regular and modified {\fentree} ({\ft}).
\label{fig:fenwick_tree}}
\end{figure}

Lastly, in Figure~\ref{fig:fenwick_tree} we show the comparison
between the regular {\fentree} and the modified version as
described above. As expected, the two curves are very similar up to
$n=2^{20}$, then the cache conflicts start to pay a significant role
in the behavior of the regular {\fentree},
making the difference between the two curves to widen progressively.
As an example, collecting the number of cache misses using Linux \textsf{perf}
for $n \approx 250 \times 10^6$
(excluding those spent during the construction
of the data structures and generation of the queries)
indeed reveals that the regular version incurs in $2.5\times$ the cache-misses
of the modified version, for both {\Sum} and {\Update}.

%

%
%

From now on, we simply refer to the modified {\fentree} \emph{without}
the problem of cache conflicts as \textsf{FT} in the plots
for further comparisons.
\section{Segment-Tree vs. Fenwick-Tree}\label{sec:segment_tree_vs_fenwick_tree}

\begin{figure}[t]
\centering

\subfloat[{\Sum}]{
\includegraphics[scale=\myfigsize]{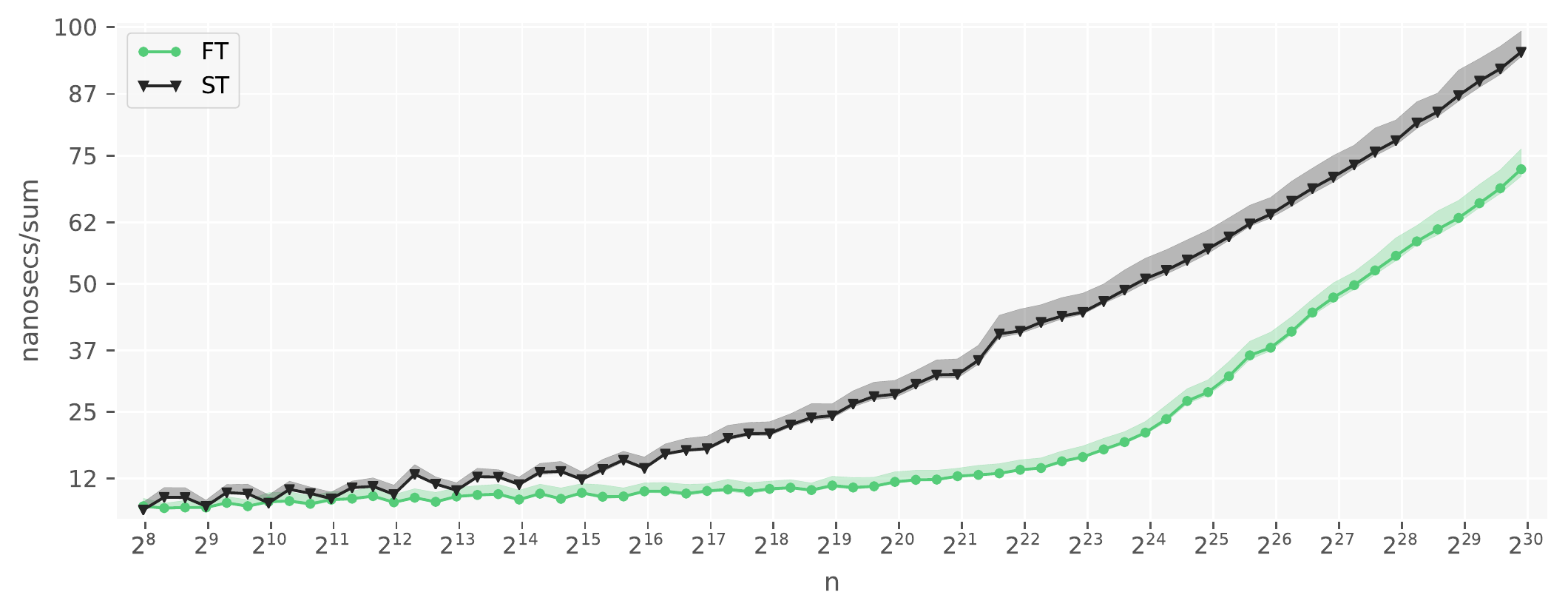}
}

\subfloat[{\Update}]{
\includegraphics[scale=\myfigsize]{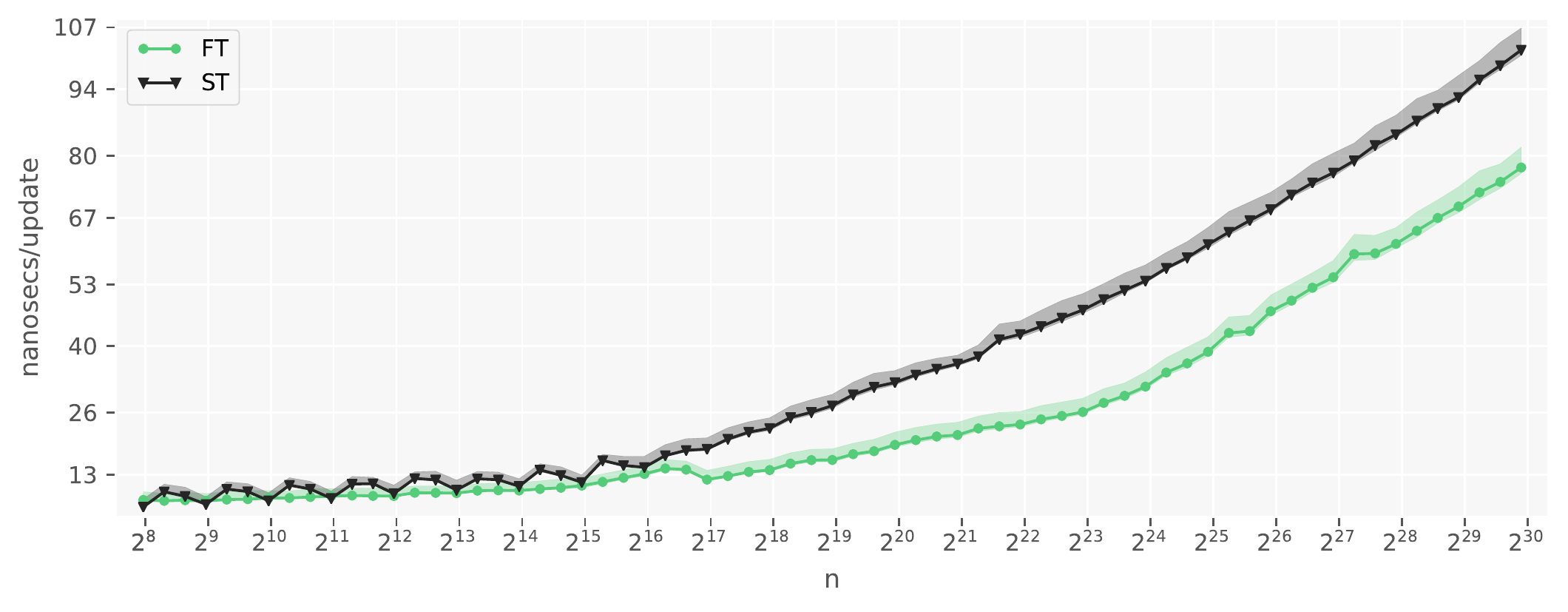}
}

\caption{The running times of {\Sum}/{\Update} for the {\segtree} (ST) and the {\fentree} ({\ft}).
\label{fig:segment_tree_vs_fenwick_tree}}
\end{figure}

In this section we compare the optimized versions
of the the {\segtree} and {\fentree} -- respectively,
the bottom-up left-sum {\segtree} with 2-loop traversal,
and the modified {\fentree} with reduced cache conflicts.
A quick look at Figure~\ref{fig:segment_tree_vs_fenwick_tree}
immediately reveals that the {\fentree}
outperforms the {\segtree}.
There are two different reasons that explain why the {\fentree}
is more efficient than the {\segtree}.

\begin{enumerate}

\item Although we have substantially simplified the code logic for traversing
the {\segtree} with the bottom-up navigation,
the code for the {\fentree} is even simpler and
requires just a few arithmetic operations
plus a memory access at each iteration of the loop.

\item While both trees have height $\lceil \log_2 n \rceil + 1$,
the number of traversed nodes per operation is different.
The {\segtree} \emph{always} traverses $\lceil \log_2 n \rceil + 1$ nodes
and that is also the number of iterations of the \texttt{while} loop.
When the branch-free code is into play,
also $\lceil \log_2 n \rceil + 1$ memory accesses are performed.
As we already explained, for the branchy implementation
the number of memory accesses
depends on the query: for random distributed
queries we expect to perform one access every two iterations of the loop
(i.e., half of the times we go left, half of the time we go right).
Now, for a random integer $i$ between 0 and $n-1$,
the number of bits set we expect to see in its binary representation
is approximately 50\%.
Therefore, the number of loop iterations \emph{and} memory accesses
the {\fentree} is likely to perform
is $\frac{1}{2} \lceil \log_2(n+1) \rceil + 1$, i.e., half of those performed
by the {\segtree}, regardless the value of $n$.

\end{enumerate}


\emph{Both} factors contribute to the difference in efficiency
between the two data structures.
However, for small values of $n$,
the number of performed instructions
is the major contributor in the running time
as both data structures fit in cache,
thus memory accesses are relatively cheap (point 1).
For example, when $n$ is $\approx 2^{16}$ and considering {\Sum} queries,
the {\fentree} code spends 58\% of the cycles spent by the {\segtree}
and executes nearly 1/3 of instructions.
It also executes half of the branches and misses 11\% of them.
Moving towards larger $n$, the memory latency progressively becomes
the dominant factor in the running time, thus favoring solutions
that spare memory accesses (point 2).
Considering again the number of cache misses
for $n \approx 250 \times 10^6$
(excluding overheads during benchmarking),
we determined that the {\fentree} incurs in 25\% less cache misses
than the {\segtree} for both {\Sum} and {\Update}.

In conclusion,
we will exclude the {\segtree} from the experiments
we are going to discuss in the rest of the paper
and compare against the {\fentree}.

%
%
%
%
%

%
%
%
%
%
%
%
\section{The $b$-ary Segment-Tree}\label{sec:segtree_bary}

The solutions analyzed in the previous sections
have two main limitations.
(1) The height of the tree is $\lceil \log_2 n \rceil + 1$:
for the {\segtree}, because each internal node has 2 children;
for the {\fentree}, because an index in $[0,n)$ is decomposed
as sum of some powers of 2.
Thus, the tree may become excessively tall for large values of $n$.
(2) The running time of {\Update} does not depend on $\Delta$.
In particular, it makes no difference whether $\Delta$ is ``small'',
e.g., it fits in one byte, or arbitrary big:
possible assumptions on $\Delta$ are not currently exploited
to achieve a better runtime.

To address these limitations, we can let each internal node of the tree
hold a block of $b>2$ keys.
While this reduces the height of the tree for
a better cache usage, it also enables the use of SIMD instructions
for faster {\Update} operations
because several keys can be updated in parallel.

Therefore, in Section~\ref{sec:nodes} we introduce a solution that
works for ``small'' arrays of $b>2$ keys, e.g., 64 or 256.
Then, in Section~\ref{sec:large_arrays}, we show how to embed this small-array
solution into the nodes of a {\segtree}
to obtain a solution for arbitrary values of $n$.


\subsection{Prefix Sums on Small Arrays: SIMD-Aware Node Layouts}\label{sec:nodes}

\begin{figure}[t]
\begin{lstlisting}
static const uint64_t all1 = uint64_t(-1);

static const uint64_t T[4 * 4] = {
    0, all1, all1, all1,
    0, 0, all1, all1,
    0, 0, 0, all1,
    0, 0, 0, 0
};

void update(uint64_t i, int64_t delta) {
    __m256i U = _mm256_set1_epi64x(delta);
    __m256i msk = _mm256_load_si256((__m256i const*)T + i);
    __m256i K = _mm256_loadu_si256((__m256i const*)keys);
    U = _mm256_and_si256(U, msk);
    _mm256_storeu_si256((__m256i*)keys, _mm256_add_epi64(U, K));
}
\end{lstlisting}
\caption{A SIMD-based implementation of {\Update}
for an array \var{keys} of 4 $\times$ 64-bit integers.
The table \var{T} must be aligned on a 32-byte boundary.
\label{alg:update4}}
\end{figure}

A reasonable starting point would be to consider $b = 4$ keys
and compute their prefix sums in an array $\var{keys}[0..4)$.
Queries are answered in the obvious way: $\Sum(i)=\var{keys}[i]$,
and we do not explicitly mention it anymore in the following.
Instead, an $\Update(i,\Delta)$ operation is solved
by setting $\var{keys}[j] = \var{keys}[j] + \Delta$ for $j = i..3$.
We can do this in parallel with SIMD as follows.
Let \var{U} be a register of $4 \times 64$ bits that packs 4 integers,
where the first $i - 1$ integers are 0 and the
remaining ones, from $j = i..3$, are equal to $\Delta$.
Then $\Update(i,\Delta)$ is achieved by adding \var{U}
and \var{keys} in parallel using the instruction
\code{\_mm256\_add\_epi64}.
To obtain the register \var{U} as desired, we
first initialize \var{U} with 4 copies of $\Delta$
using the instruction \code{\_mm256\_set1\_epi64x}.
Now, the copies of $\Delta$ before the $i$-th must be masked out.
To do so, we use the index $i$ to perform a lookup into
a small pre-computed table \var{T}, where $\var{T}[i]$ is the
proper 256-bit mask.
In this case, \var{T} is a $4 \times 4$ table of unsigned 64-bit integers,
where
$\var{T}[0][0..3] = [0,2^{64}-1,2^{64}-1,2^{64}-1]$,
$\var{T}[1][0..3] = [0,0,2^{64}-1,2^{64}-1]$,
$\var{T}[2][0..3] = [0,0,0,2^{64}-1]$, and
$\var{T}[3][0..3] = [0,0,0,0]$.
Once we loaded the proper mask,
we obtain the wanted register configuration with
$\var{U} = \code{\_mm256\_and\_si256}(\var{U},\var{T}[i])$.
The code\footnote{For all the code listings we show in this section,
it is assumed that the size in bytes of
(\code{u})\code{int64\_t}, (\code{u})\code{int32\_t}, and (\code{u})\code{int16\_t}
is 8, 4, and 2, respectively.}
for this algorithm is shown in Figure~\ref{alg:update4}.



An important observation is in order, before proceeding.
One could be tempted to leave the integers in $\var{keys}[0..4)$
as they are in order to obtain trivial updates
and use SIMD instructions to answer {\Sum}.
During our experiments we determined that this solution gives
a \emph{worse} trade-off than the one described above: this was
actually no surprise,
considering that the algorithm for computing prefix sums with SIMD
is complicated as it involves several shifts and additions (besides
load and store).
Therefore, SIMD is more effective on updates rather than queries.

\begin{figure}[t]
\centering
\includegraphics[scale=1
]{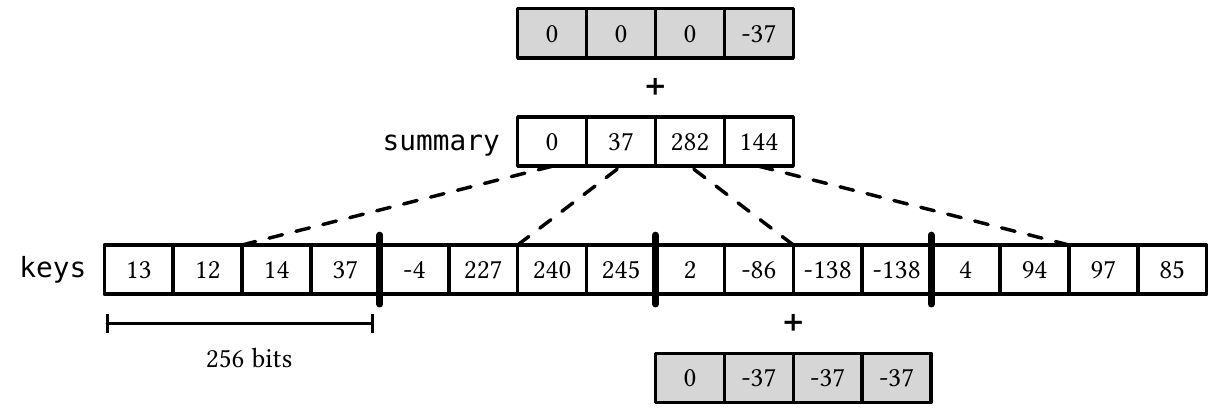}
\caption{A two-level node layout built from the array
$[$13, -1, 2, 23, -4, 231, 13, 5, 2, -88, -52, 0, 4, 90, 3, -12$]$.
The shaded arrays represent the content of the registers
used when executing the operation $\Update(9,-37)$.
\label{fig:node16}}
\end{figure}

\paragraph*{Two-Level Data Structure}
As already motivated, we would like to consider larger values of $b$,
e.g., 64 or 256, in order to obtain even flatter trees.
Working with such branching factors, would mean to apply
the algorithm in Figure~\ref{alg:update4}
for $b/4$ times, which may be too slow.
To alleviate this issue, we use a two-level data structure.
We split $b$ into $\sqrt{b}$ segments and store each segment
in prefix-sum. The sum of the integers in the $j$-th segment
is stored in a \var{summary} array of $\sqrt{b}$ values
in position $j + 1$, for $j = 0..\sqrt{b}-2$
($\var{summary}[0] = 0$).
The values of the \var{summary} are stored in prefix-sum as well.
In Figure~\ref{fig:node16} we show a graphical example of this
organization for $b=16$.
In the example, we apply the algorithm in Figure~\ref{alg:update4}
only twice (first on the \code{summary}, then on a specific
segment of \code{keys}): without the two-level organization,
4 executions of the algorithm would have been needed.
Instead, queries are as easy as
$\Sum(i) = \var{summary}[i / \sqrt{b}] + \var{keys}[i]$.



The code corresponding to this approach for $b=64$
is shown in Figure~\ref{alg:update8}\footnote{The \code{build} method of the \code{node} class
builds the two-level data structure as we explained above, and writes
\code{node::size} bytes onto the output buffer, \code{out}.
We do not report it for conciseness.}.
In this case, note that the whole \var{summary} fits
in one cache line,
because its size is $8 \times 8 = 64$ bytes,
as well as each segment of \var{keys}.
The table \var{T} stores $(8 + 1) \times 8$ 64-bit unsigned values,
where $\var{T}[i][0..7]$ is a an array of 8 integers: the first $i$ are 0
and the other $8-i$ are $2^{64}-1$, for $i=0..8$.

Lastly, the space overhead due to the \var{summary} array
is always $1/\sqrt{b} \times 100\%$.
For the example code in Figure~\ref{alg:update8}, the space consumed
is 12.5\% more than that of the input array (576 bytes consumed
instead of $64 \times 8 = 512$).

\begin{figure}[t]
\begin{lstlisting}
struct node {
    static const uint64_t fanout = 64, size = 576;

    node(uint8_t* ptr)
        : summary(reinterpret_cast<int64_t*>(ptr))
        , keys(reinterpret_cast<int64_t*>(ptr) + 8)
    {}

    static void build(int64_t const* input, uint8_t* out);

    void update(uint64_t i, int64_t delta) {
        uint64_t j = i / 8, k = i % 8;
        __m256i U = _mm256_set1_epi64x(delta);

        __m256i msk_j0 = _mm256_load_si256((__m256i const*)(T + 8) + 2 * j + 0);
        __m256i msk_j1 = _mm256_load_si256((__m256i const*)(T + 8) + 2 * j + 1);
        __m256i U_j0 = _mm256_and_si256(U, msk_j0);
        __m256i U_j1 = _mm256_and_si256(U, msk_j1);
        __m256i dst_summary0 = _mm256_loadu_si256((__m256i const*)summary + 0);
        __m256i dst_summary1 = _mm256_loadu_si256((__m256i const*)summary + 1);
        _mm256_storeu_si256((__m256i*)summary + 0, _mm256_add_epi64(U_j0, dst_summary0));
        _mm256_storeu_si256((__m256i*)summary + 1, _mm256_add_epi64(U_j1, dst_summary1));

        __m256i msk_k0 = _mm256_load_si256((__m256i const*)T + 2 * k + 0);
        __m256i msk_k1 = _mm256_load_si256((__m256i const*)T + 2 * k + 1);
        __m256i U_k0 = _mm256_and_si256(U, msk_k0);
        __m256i U_k1 = _mm256_and_si256(U, msk_k1);
        __m256i dst_keys0 = _mm256_loadu_si256((__m256i const*)(keys + j * 8) + 0);
        __m256i dst_keys1 = _mm256_loadu_si256((__m256i const*)(keys + j * 8) + 1);
        _mm256_storeu_si256((__m256i*)(keys + j * 8) + 0, _mm256_add_epi64(U_k0, dst_keys0));
        _mm256_storeu_si256((__m256i*)(keys + j * 8) + 1, _mm256_add_epi64(U_k1, dst_keys1));
    }

    int64_t sum(uint64_t i) const { return summary[i / 8] + keys[i]; }

private:
    int64_t *summary, *keys;
};
\end{lstlisting}
\caption{Code for a two-level node data structure with $b=64$.
\label{alg:update8}}
\end{figure}

\begin{figure}[t]
\begin{lstlisting}
struct node {
    static const uint64_t fanout = 64, size = 721;

    node(uint8_t* ptr)
        : summary(reinterpret_cast<int64_t*>(ptr))
        , keys(reinterpret_cast<int64_t*>(ptr) + 8)
        , summary_buffer(reinterpret_cast<int16_t*>(ptr) + 288)
        , keys_buffer(reinterpret_cast<int16_t*>(ptr) + 296)
        , updates(ptr + size - 1)
    {}

    static void build(int64_t const* input, uint8_t* out);

    void update(uint64_t i, int8_t delta) {
        uint64_t j = i / 8, k = i % 8;
        __m128i U = _mm_set1_epi16(delta);

        __m128i msk_j = _mm_load_si128((__m128i const*)(T + 8) + j);
        __m128i U_j = _mm_and_si128(U, msk_j);
        __m128i dst_summary_buffer = _mm_loadu_si128((__m128i const*)summary_buffer);
        _mm_storeu_si128((__m128i*)summary_buffer, _mm_add_epi16(U_j, dst_summary_buffer));

        __m128i msk_k = _mm_load_si128((__m128i const*)T + k);
        __m128i U_k = _mm_and_si128(U, msk_k);
        __m128i dst_keys_buffer = _mm_loadu_si128((__m128i const*)keys_buffer + j);
        _mm_storeu_si128((__m128i*)keys_buffer + j, _mm_add_epi16(U_k, dst_keys_buffer));

        *updates += 1;
        if (*updates == 255) {
            for (uint64_t l = 0; l != 8; ++l) {
                summary[l] += summary_buffer[l];
                summary_buffer[l] = 0;
            }
            for (uint64_t l = 0; l != 64; ++l) {
                keys[l] += keys_buffer[l];
                keys_buffer[l] = 0;
            }
        }
    }

    int64_t sum(uint64_t i) const {
        return summary[i / 8] + keys[i] + summary_buffer[i / 8] + keys_buffer[i];
    }

private:
    int64_t *summary, *keys; int16_t *summary_buffer, *keys_buffer; uint8_t *updates;
};
\end{lstlisting}
\caption{Code for a two-level node data structure with $b=64$, for
the ``restricted'' case where $\Delta$ is a signed 8-bit integer.
\label{alg:update8_restricted}}
\end{figure}

\paragraph*{Handling Small Updates}
It is intuitive that one can obtain faster results for {\Update}
if the bit-width of $\Delta$ is smaller than 64.
In fact, a restriction on the possible values $\Delta$ can take
permits to ``pack'' more such values
inside a SIMD register which, in turn,
allows to update a larger number of keys in parallel.
As we already observed, neither the binary {\segtree} nor
the {\fentree} can possibly take advantage of this restriction.

To exploit this possibility, we buffer the updates.
We restrict the bit-width of $\Delta$ to 8, that is,
$\Delta$ is now a value in the range $[-128, +127]$
(instead of the generic, un-restricted, case of $\Delta \in [-2^{63},+2^{63}-1]$).
We enrich the two-level node layout introduced before
with some additional arrays to buffer the updates.
These arrays are made of 16-bit signed integers.
We maintain one such array of size $\sqrt{b}$, say \var{summary\_buffer},
plus another of size $b$, \var{keys\_buffer}.
In particular, the $i$-th value of \var{keys\_buffer}
holds the $\Delta$ value for the $i$-th key;
similarly, the $(j+1)$-th value of \var{summary\_buffer}
holds the $\Delta$ value for the $j$-th segment.
The buffers are kept in prefix-sum.

Upon an $\Update(i,\Delta)$ operation, the buffer for the summary
and that of the specific segment comprising $i$
are updated using SIMD.
The key difference now is that -- because we work with smaller
integers -- 8, or even 16, integers are updated simultaneously,
instead of only 4 as illustrated with the code in Figure~\ref{alg:update4}.

For example, suppose $b=64$.
The whole \var{summary\_buffer},
which consists of $16 \times 8 = 128$ bits,
fits into one SSE SIMD register,
thus 8 integers are updated simultaneously.
For $b=256$, 16 integers are updated simultaneously
because $16 \times 16 = 256$ bits again fit into one AVX SIMD register.
This makes a big improvement with respect to the un-restricted
case because instead of executing the algorithm in Figure~\ref{alg:update4}
for $\sqrt{b}/4$ times, \emph{only one} such update is sufficient.
This potentially makes the restricted case
$2\times$ and $4\times$ faster than the un-restricted case
for $b=64$ and $b=256$ respectively.

To avoid overflow issues,
we bring the \var{keys} (and the \var{summary}) up-to-date
by reflecting on these the updates stored in the buffer.
Since we can perform a maximum of $-2^{15}/-128 = 256$
updates before overflowing, we clean the buffers
every 256 {\Update} operations.
The solution described here holds, therefore, in the
amortized sense\footnote{Note that such ``cleaning'' operation will become
less and less frequent the deeper a node in the tree
hierarchy. Thus, scalar code is efficient to perform this operation
as vectorization would negligibly affect the running time.
}.

Figure~\ref{alg:update8_restricted} contains the relevant
code illustrating this approach.
The code should be read as the ``restricted'' variant
of that shown in Figure~\ref{alg:update8}.
We count the number of updates with a single
8-bit unsigned integer, \var{updates},
which is initialized to 255 in the \code{build} method.
When this variable is equal to 255, it will overflow
the next time it is incremented by 1,
indeed making it equal to 0.
Therefore we correctly clean the buffers every 256 updates.

In this case, the table \var{T} stores $(16+1) \times 16$
16-bit unsigned values, where each $\var{T}[i][0..15]$ contains a prefix
of $i$ zeros, followed by $16-i$ copies of $2^{16}-1$,
for $i=0..16$.

Also, {\Sum} queries are now answered by computing the sum
between four quantities, which is actually more expensive than
the queries for the general case.

The data structure consumes
$(2b + 10\sqrt{b} + 1)/(8b) \times 100\%$ more bytes than
the input. For $b=64$, as in the code given in Figure~\ref{alg:update8_restricted},
this extra space is 40.8\%; for $b=256$, it is 32.9\%.

\begin{figure}[t]
\centering
\includegraphics[scale=1]{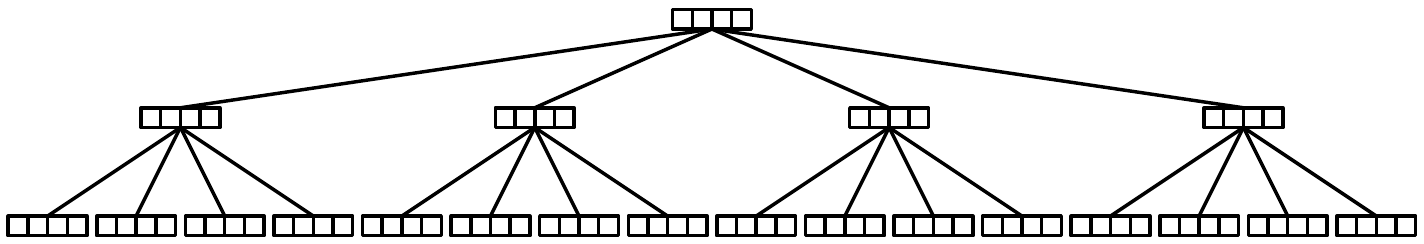}
\caption{The shape of $b$-ary {\segtree} with $b=4$,
built from an input array of size 64.
\label{fig:segtree_shape}}
\end{figure}

\subsection{Prefix Sums on Large Arrays}\label{sec:large_arrays}

In this section we use
the two-level data structures introduced in Section~\ref{sec:nodes}
in the nodes of a {\segtree}, to establish new practical trade-offs.
If $b$ is the node's fanout, this solution provides
a tree of height $\lceil \log_b n \rceil$,
whose shape is illustrated in Figure~\ref{fig:segtree_shape}
for the case $b=4$ and $n=64$.
The code in Figure~\ref{code:simdsegtree}
takes a generic \code{Node} structure as a template parameter and
builds the tree based on its fanout and size (in bytes).
In what follows, we shall first discuss some optimizations and then
illustrate the experimental results.


\paragraph*{Avoiding Loops and Branches}
The tree data structure is serialized in an array
of bytes, the \code{tree} array in the code.
In a separate array, \code{nodes}, we keep instead the number of nodes
at each level of the tree.
This information is used to traverse the tree.
In particular, when we have to resolve an operation
at a given tree level, all we have to do is to
instantiate a \code{Node} object
at a proper byte location in the array \code{tree}
and call the desired {\Sum}/{\Update} method of the object.
Such proper byte location depends on both the number of nodes
in the previous levels and the size (in bytes)
of the used \code{Node} structure.

The deliberative choice of working with large node fanouts,
such as 64 and 256, makes the tree very flat.
For example when $b=256$, the {\segtree} will be of
height at most 3 until $n$ is $2^{24}$,
and actually only at most 4 for arrays of up to $2^{32}$ keys.
This suggests to write a specialized code path
that handles each possible value of the tree height.
Doing so permits to completely eliminate the loops
in the body of both {\Sum} and {\Update} (unrolling)
and reduce possible data dependencies,
taking into account that the result of an operation
at a given level of the tree
does \emph{not} depend on the
result of the operations at the other levels.
Therefore, instead of looping through the levels of the tree, like


\begin{lstlisting}
int64_t sum(uint64_t i) const
    int64_t sum = 0;
    for (uint32_t h = 1; h != Height; ++h) {
        Node node(...);
        sum += node.sum(...);
    }
    return sum;
}
\end{lstlisting}
that actually stalls the computation at level \var{h} because that
at level \code{h+1} cannot begin,
we instantiate a different \code{Node} object for each level without a loop.
For example if the height of the tree is 2, we do
\begin{lstlisting}
uint64_t child1 = i / b;
Node node1(ptr);
Node node2(ptr + (1 + child1) * Node::size);
/* do something with node1 and node2 */
\end{lstlisting}
and if it is 3, we do instead
\begin{lstlisting}
uint64_t child1 =    i / (b * b);
uint64_t child2 = (i % (b * b)) / b;
Node node1(ptr);
Node node2(ptr + (1 + child1) * Node::size);
Node node3(ptr + (1 + nodes[1] + child2 + child1 * b) * Node::size);
/* do something with node1, node2, and node3 */
\end{lstlisting}
where \var{ptr} is the pointer to the memory holding the \var{tree} data.

Lastly, we discuss why also the height of the tree, \var{Height} in the code,
is modeled as a template parameter.
If the value of \var{Height} is known at compile-time, the compiler
can produce a template specialization of the \var{segment\_tree} class
that avoids the evaluation of an \code{if-else}
cascade that would have been otherwise necessary to select
the proper code-path that handles that specific value of \code{Height}.
This removes unnecessary branches in the code of the operations,
and it is achieved with the \code{if-constexpr} idiom of C++17:
\begin{lstlisting}
int64_t sum(uint64_t i) const {
    if constexpr (Height == 1) { ... }
    if constexpr (Height == 2) { ... }
    ...
}

void update(uint64_t i, int64_t delta) {
    if constexpr (Height == 1) { ... }
    if constexpr (Height == 2) { ... }
    ...
}
\end{lstlisting}
where the code inside each \code{if} branch handles the corresponding
value of \code{Height} as we explained above.

\begin{figure}[t]
\begin{lstlisting}
template <uint32_t Height, typename Node>
struct segment_tree_bary {
    static_assert(Height > 0);
    static const uint64_t b = Node::fanout;

    void build(int64_t const* input, uint64_t n) {
        uint64_t m = n, total_nodes = 0;
        nodes.resize(Height);
        for (int h = Height - 1; h >= 0; --h) {
            m = ceil(static_cast<double>(m) / b);
            nodes[h] = m;
	    total_nodes += m;
        }

        assert(m == 1);
        uint64_t tree_size = total_nodes * Node::size + 4 * Height;
        tree.resize(tree_size);
	ptr = tree.data();

        vector<int64_t> node(b);
	uint8_t* begin = tree.data() + tree_size;
        uint64_t step = 1;
        for (int h = Height - 1; h >= 0; --h, step *= b) {
            begin -= nodes[h] * Node::size;
	    uint8_t* out = begin;
            for (uint64_t i = 0, base = 0; i != nodes[h]; ++i) {
                for (int64_t& x : node) {
                    int64_t sum = 0;
                    for (uint64_t k = 0; k != step and base + k < n; ++k)
			sum += input[base + k];
                    x = sum;
		    base += step;
                }
                Node::build(node.data(), out);
		out += Node::size;
            }
        }
    }

    int64_t sum(uint64_t i) const;
    void update(uint64_t i, int64_t delta);

private:
    uint8_t* ptr;
    vector<uint32_t> nodes;
    vector<uint8_t> tree;
};
\end{lstlisting}
\caption{The $b$-ary {\segtree} code handling a \code{Node}
structure that is specified as a template parameter.
\label{code:simdsegtree}}
\end{figure}

\begin{figure}[t]
\centering

\subfloat[{\Sum}]{
\includegraphics[scale=\myfigsize]{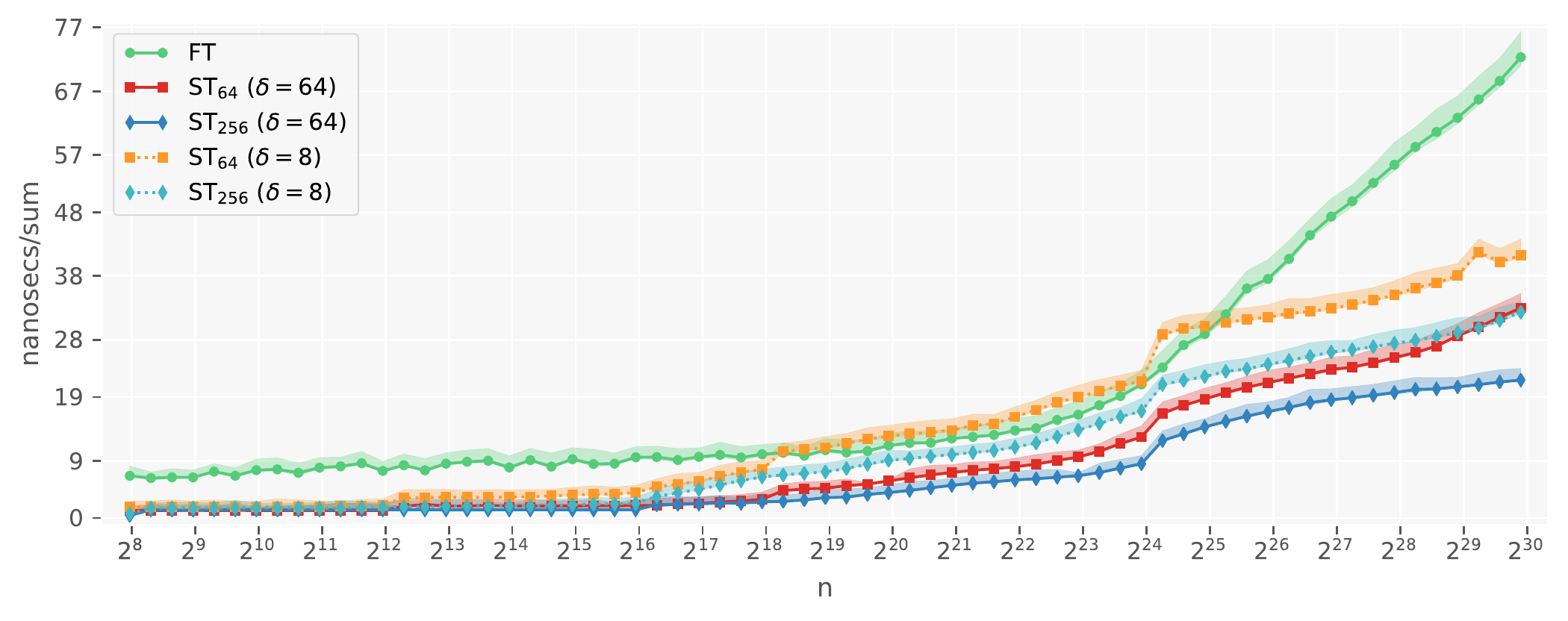}
\label{fig:segment_tree_a}
}

\subfloat[{\Update}]{
\includegraphics[scale=\myfigsize]{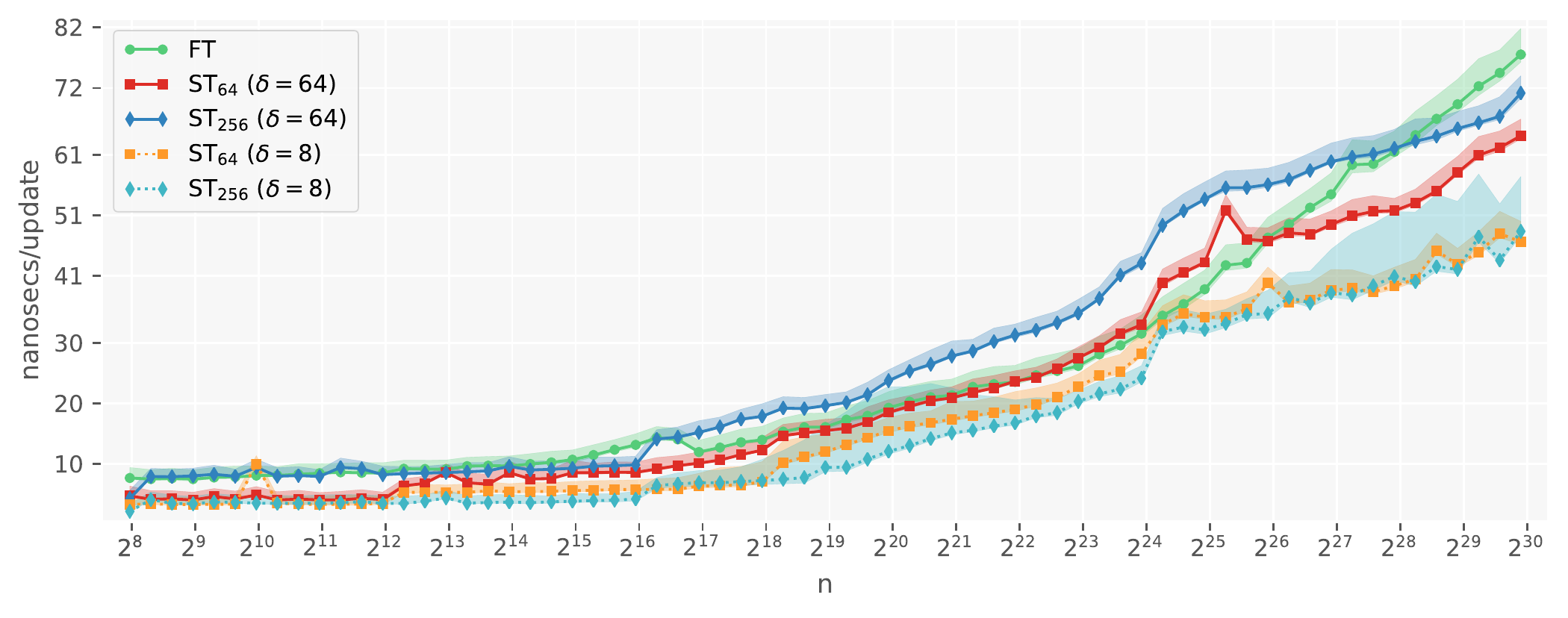}
\label{fig:segment_tree_b}
}

\caption{The running times of {\Sum}/{\Update}
for the $b$-ary {\segtree} with branching factor 64 and 256.
\label{fig:segment_tree}}
\end{figure}

\paragraph*{Experimental Results}
We now comment on the experimental results in Figure~\ref{fig:segment_tree}.
The two version of the data structure that we tested, with
$b=64$ and $b=256$, will be referred
to as ${\sst}_{64}$ and ${\sst}_{256}$ respectively.


We first consider the most general case where $\delta=64$.
Regarding {\Sum} queries,
we see that the two-level node layout, which guarantees
constant-time complexity, together with the very short tree height,
makes both the tested {\segtree} versions substantially
improve over the {\fentree}.
The curves for the {\segtree} are strictly sandwiched between 1 and 2 nanoseconds
until the data structures cannot be contained in $L_2$, which happens
close to the $2^{17}$ boundary. In this region, the {\segtree}
is bound by the CPU cycles,
whereas the {\fentree} by branch misprediction.
For $n \approx 2^{16}$ and {\Sum} queries,
the ${\sst}_{64}$ executes 23\% of the instructions and 22\% of the
cycles of the {\fentree}.
As a result of the optimizations we discussed at the beginning of the section
(loop-unrolling and reduced branches),
it only performs 2.2\% of the branches of the {\fentree} and misses 0.36\% of those.
The larger tree height of the {\fentree} also induces a poor cache utilization
compared to the {\segtree}. As already discussed,
this becomes evident for large values of $n$.
For example, for $n=250 \times 10^6$ and {\Sum} queries,
${\sst}_{64}$ incurs in 40\% less cache misses than the {\fentree}
(and ${\sst}_{256}$ in 80\% less).
This cache-friendly behavior is a direct consequence of the very short height.

%
%
%

%

\begin{figure}[t]
\centering
\includegraphics[scale=\myfigsize]{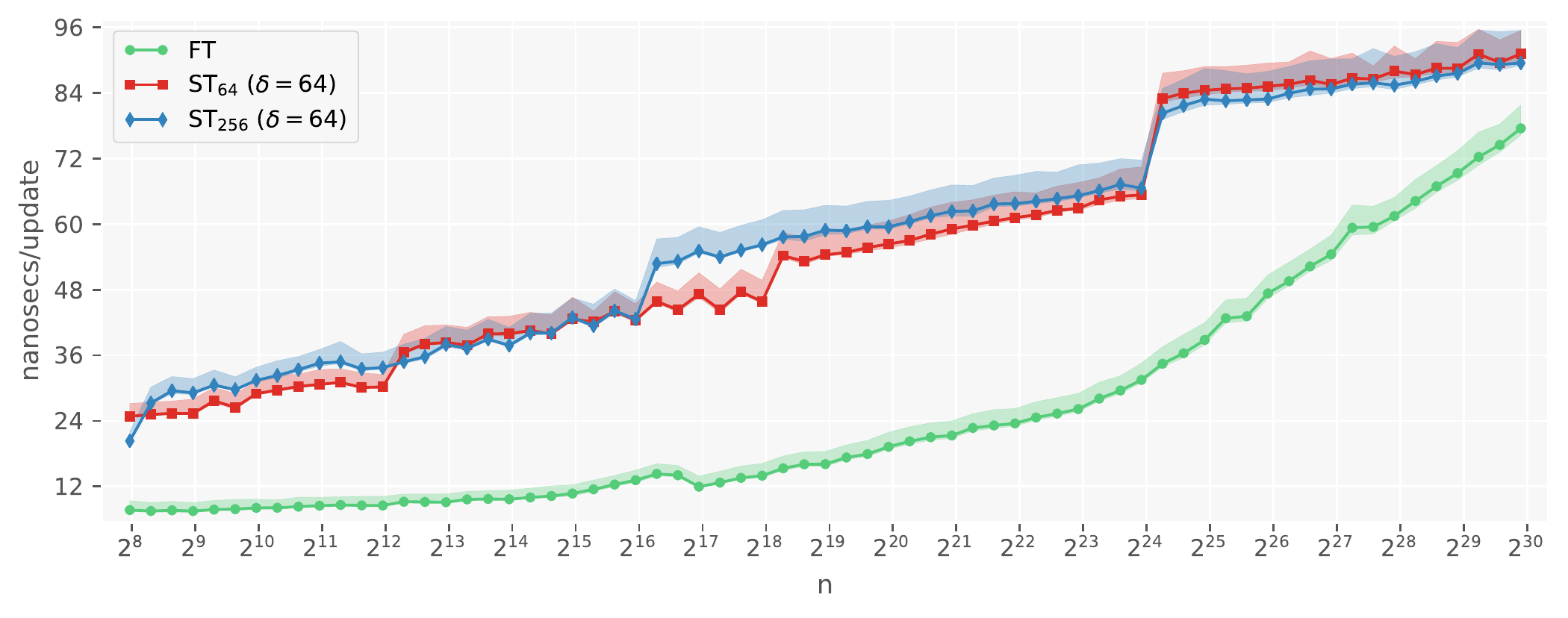}
\caption{The running times of {\Update}
for the $b$-ary {\segtree} with branching factor 64 and 256,
\emph{without} manual usage of SIMD instructions.
\label{fig:segment_tree_no_simd}}
\end{figure}

Perhaps more interesting is the case for updates, where SIMD instructions
can be exploited to lower the running time.
As a reference point, we show in Figure~\ref{fig:segment_tree_no_simd}
the running time of {\Update}
\emph{without} manual usage of SIMD (only the general
case is shown for simplicity): compared to the plots in Figure~\ref{fig:segment_tree},
we see that SIMD offers a good reduction
in the running time of {\Update}.
If we were \emph{not} to use SIMD,
we would have obtained a $4\times$ lager time for {\Update}, thus SIMD
is close to its ideal speed-up for our case
(4 $\times$ 64-bit integers are packed in a register).
This reduction is more
evident for $b=64$ rather than $b=256$ because, as we discussed
in Section~\ref{sec:nodes}, updating 8 keys is faster than
updating 16 keys: this let the $\sst_{64}$ perform (roughly)
half of the SIMD instructions of $\sst_{256}$,
although $\sst_{256}$ is one level shorter.
This translates into fewer spent cycles and less time.
Also, ${\sst}_{256}$ incurs in a bit more cache misses than ${\sst}_{64}$,
11\% more,
because each {\Update} must access the double of cache lines
than ${\sst}_{64}$: from Section~\ref{sec:nodes}, recall that
each segment consists of 16 keys that fit in two cache lines.
Instead, the $\sst_{64}$ solution is as fast or better
than the {\fentree} thanks to SIMD instructions.
Is is important to stress that $\sst_{64}$ actually executes
more instructions then the {\fentree}. However, it does so
in fewer cycles, thus confirming the good impact of SIMD.

Considering the overall trade-off between {\Sum} and {\Update},
we conclude that -- for the general case where $\Delta$ is expressed
in 8 bytes -- the ${\sst}_{64}$ data structure offers the best running times.

We now discuss the case where $\delta = 8$.
Queries are less efficient than those in the general case because, as explained
in Section~\ref{sec:nodes}, they are computed by accessing four different
array locations at each level of the tree.
These additional accesses to memory induce
more cache-misses as soon as the data structure does not fit in cache: in fact,
notice how the time rises around $2^{12}$ and $2^{16}$ boundaries,
also because the restricted versions consume more space.
Anyway, the $\sst_{256}$ variant maintains a good advantage against the
{\fentree},
especially on the large values of $n$ (its curve is actually very similar to that of
${\sst}_{64}$ for the general case).

The restriction on $\Delta$ produces an improvement for {\Update}
as expected. Note how the shape of $\sst_{256}$ dramatically improves
(by $1.5-2\times$)
in this case because 16 keys are updated in parallel rather than just 4
as in the general case (now, 16 $\times$ 16-bit integers are packed in a register).

Therefore, we conclude that -- for the restricted case
where $\Delta$ is expressed
in 1 byte --
the solution $\sst_{256}$ offers the best trade-off between the running times
of {\Sum} and {\Update}.

%
%
%

%
%
%
%
%

\begin{figure}[t]
\centering

\subfloat[$b$-ary {\fentree}]{
\includegraphics[scale=1]{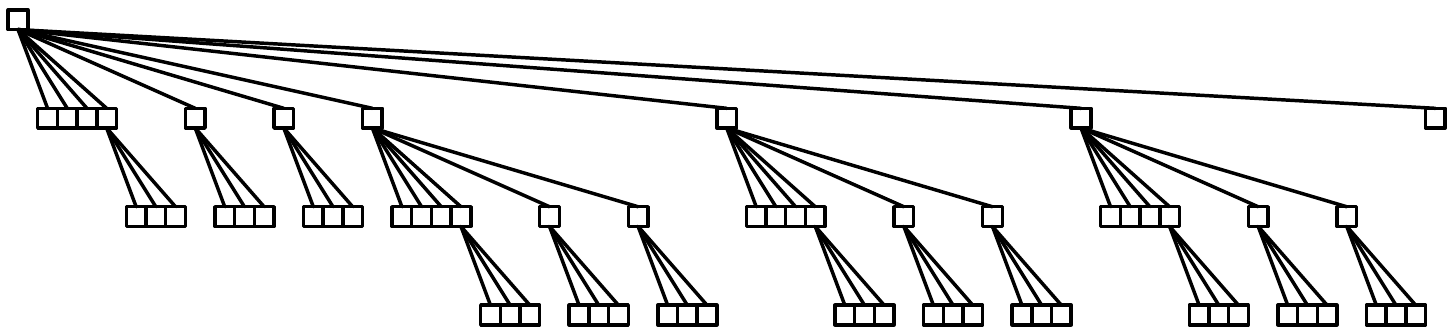}
\label{fig:bary_fentree_shape}
}

\subfloat[blocked {\fentree}]{
\includegraphics[scale=1]{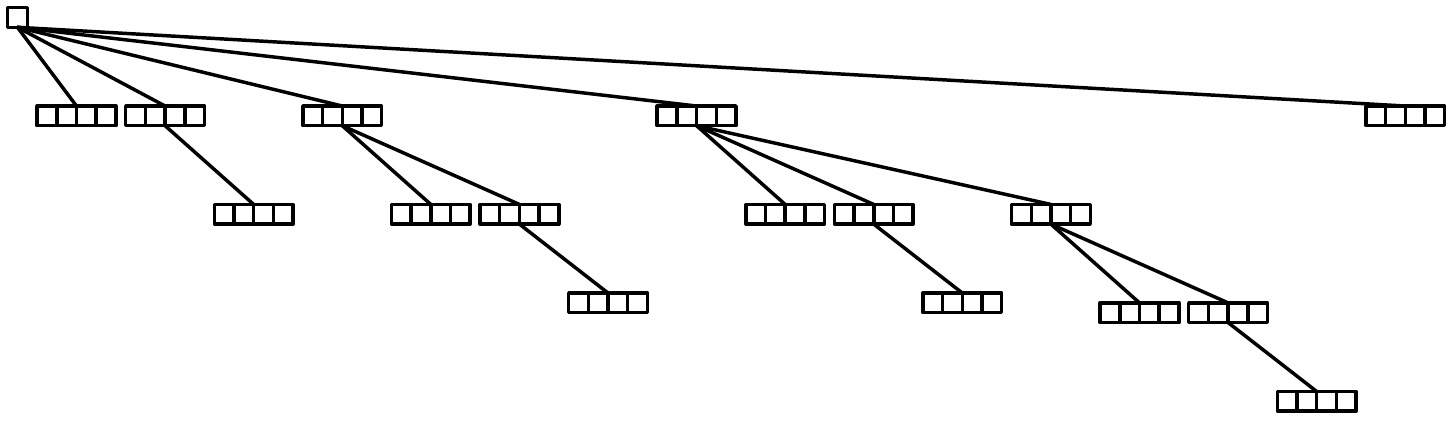}
\label{fig:blocked_fentree_shape}
}

\subfloat[truncated {\fentree}]{
\includegraphics[scale=1]{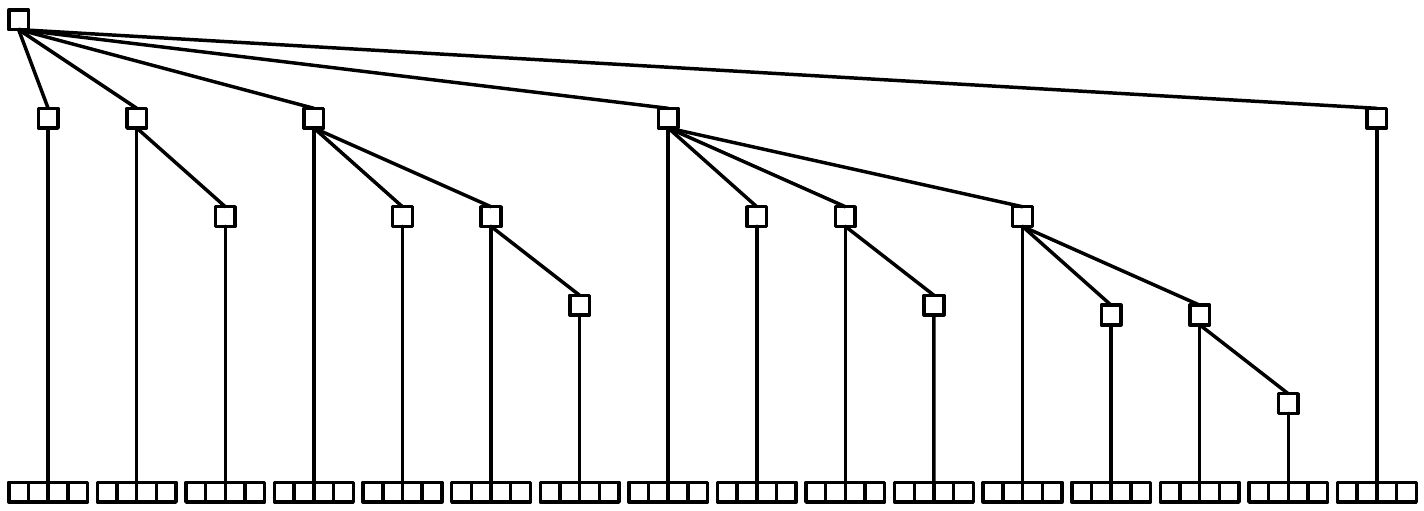}
\label{fig:truncated_fentree_shape}
}

\caption{The shapes of different data structures with $b=4$,
built from an input array of size 64,
and based on the {\fentree}.
\label{fig:fentree_data_structures}}
\end{figure}

\section{The $b$-ary Fenwick-Tree}\label{sec:fentree_bary}


The classic {\fentree} we described in Section~\ref{sec:fentree}
exploits the base-2 representation of a number in $[0,n)$
to support {\Sum} and {\Update} in time $O(\log n)$.
If we change the base of the representation to a value $b>2$,
the corresponding $b$-ary {\fentree} can be defined~\cite{bille2017succinct} --
a data structure supporting {\Sum} in $O(\log_b n)$
and {\Update} in $O(b\log_b n)$.
A pictorial representation of the data structure is given in
Figure~\ref{fig:bary_fentree_shape} for $b=4$ and $n=64$.
However,
this data structure does not expose an improved trade-off compared
to the solutions described in the previous sections, for
the reasons we discuss in the following.


Let us consider {\Sum}.
Recall that the {\fentree} ignores digits that are 0.
We have already commented on this for the case $b=2$
in Section~\ref{sec:segment_tree_vs_fenwick_tree}: for random queries,
this gives a consistent boost over the {\segtree} with $b=2$ because roughly 50\%
of the levels are skipped, \emph{as if} the height of the tree were actually
$\frac{1}{2}\lceil\log_2(n+1)\rceil$.
Unfortunately, this advantage does \emph{not} carry over for larger $b$
because the probability that a digit is 0 is $1/b$ which is very low
for the values of $b$ we consider in our experimental analysis (64 and 256).
In fact, the $b$-ary {\fentree} is not faster than the
$b$-ary {\segtree} (although it is when compared
to the classic {\fentree} with $b=2$).

Even more problematic is the case for {\Update}.
Having to deal with more digits clearly
slows down {\Update} that needs to access $b-1$
nodes per level, for a complexity of $O(b\log_b n)$.
Observe that $(b-1)\frac{\log_2 n}{\log_2 b}$ is more than $\log_2 n$
for every $b>2$. For example, for $b=64$ we can expect a slowdown
of more than $10\times$ (and nearly $32\times$ for $b=256$).
Experimental results confirmed this analysis.
Note that the $b$-ary {\segtree} does much better than this because:
(1) it traverses $\lceil \log_b n \rceil$ nodes for operation,
(2) the $b$ keys to update per node are contiguous
in memory for a better cache usage and, hence,
(3) are amenable to the SIMD optimizations
we described in Section~\ref{sec:nodes}.
Therefore, we experimented with two other ideas
in order to improve the trade-off of the $b$-ary {\fentree}.

\begin{figure}[t]
\centering

\subfloat[{\Sum}]{
\includegraphics[scale=\myfigsize]{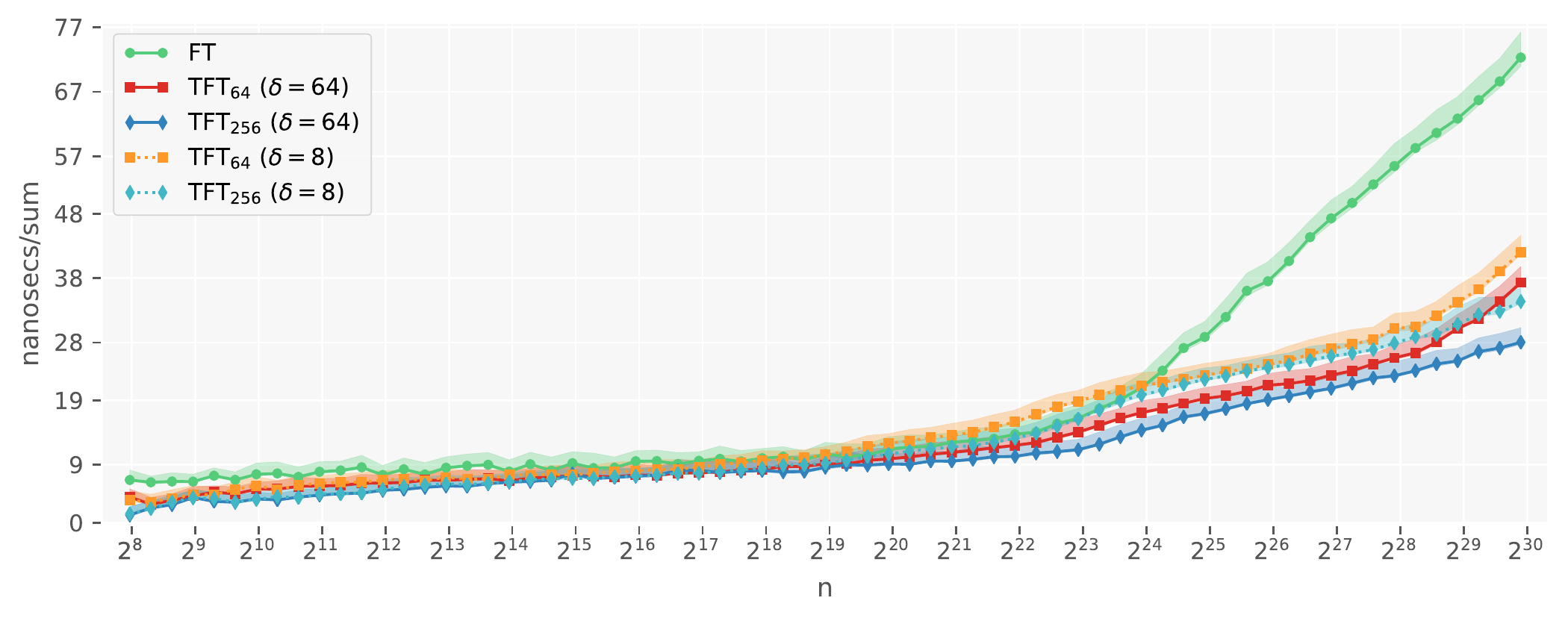}
\label{fig:fenwick_tree_truncated_a}
}

\subfloat[{\Update}]{
\includegraphics[scale=\myfigsize]{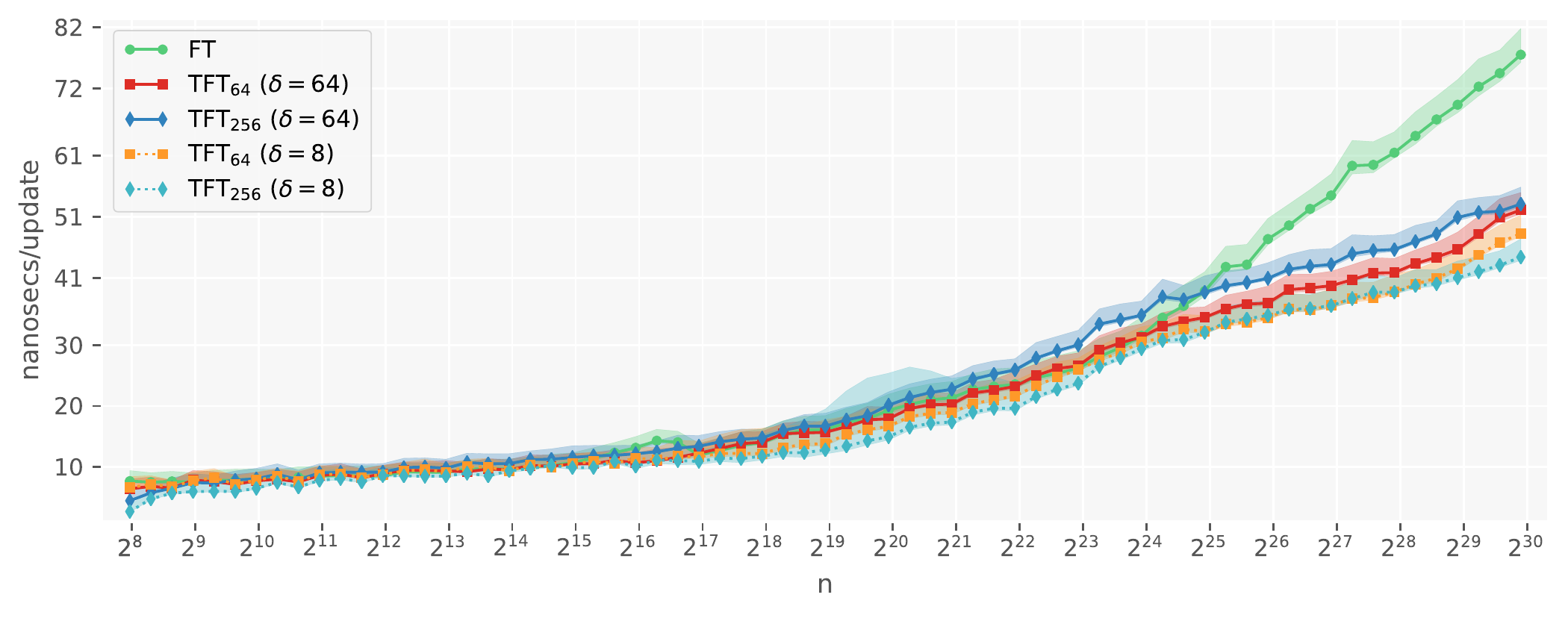}
\label{fig:fenwick_tree_truncated_b}
}

\caption{The running times of {\Sum}/{\Update}
for the truncated {\fentree} (\textsf{TFT}) with leaves of 64 and 256 keys.
\label{fig:fenwick_tree_truncated}}
\end{figure}

The first idea
is to block $b$ keys together into the nodes
of a classic {\fentree}, as depicted in Figure~\ref{fig:blocked_fentree_shape}
for $b=4$ and $n=64$.
Compared to the $b$-ary {\fentree}, this variant
-- that we call the \emph{blocked} {\fentree} --
slows down the queries but improves the updates,
for a better trade-off.
However, it does not improve over the classic
{\fentree} with $b=2$.
In fact, although the height of the tree is
$\lceil \log_2((n+1)/b) \rceil + 1$, thus smaller than that
of the classic {\fentree} by $\log_2 b$,
the number of traversed nodes is only reduced by $\frac{1}{2}\log_2 b$, which is
quite small for the values of $b$ considered (e.g., just 3 for $b=64$).
Therefore, this variant traverses slightly fewer nodes but the computation
at each node is more expensive (more cache lines accessed due to two-level nodes;
more cycles spent due to SIMD), resulting in a larger runtime
compared to the classic {\fentree}.

This suggests the implementation of a strategy
where one keeps running the simplest code for {\Update},
e.g., that of the classic {\fentree}, until the number of leaves
in the sub-tree is $b$ and, thus, these can be updated in parallel with SIMD.
The shape of the resulting data structure
-- the \emph{truncated} {\fentree} (\textsf{TFT}) --
is illustrated in Figure~\ref{fig:truncated_fentree_shape} (for $b=4$ and $n=64$)
and shows the clear
division between the upper part represented by the {\fentree} and
the lower part that consists in an array of blocks.
Compared to the classic {\fentree},
it is now intuitive that this variant reduces the number of cache-misses
because the upper part is likely to fit well in cache and $b$ keys inside
a block are contiguous in memory.

The experimental results shown in Figure~\ref{fig:fenwick_tree_truncated}
meet our expectations, as the truncated variant actually improves over
the classic {\fentree}.
In particular, it performs similarly to the {\fentree} for
small values of $n$ but gains a significant advantage for larger $n$
thanks to its better cache usage.
Comparing Figure~\ref{fig:fenwick_tree_truncated_b} to Figure~\ref{fig:segment_tree_b}
at page~\pageref{fig:segment_tree_b},
we can also observe that this variant improves over the $b$-ary {\segtree}
in the general case for {\Update} ($\delta=64$)
and large $n$ because the use of SIMD is limited
to one block of $b$ keys per operation.
For {\Sum} queries instead, the data structure performs similarly to the
$b$-ary {\segtree}, with the latter being generally faster for all values of $n$.

As last note, we remark that all the variants of the {\fentree}
we sketched in this section
are anyway part of the software library available at
\url{https://github.com/jermp/psds}.

\section{Conclusions and future work}\label{sec:conclusions}

We described, implemented, and studied the practical performance of several
tree-shaped data structures to solve the \emph{prefix-sum problem}.
After a careful experimental analysis,
the following take-away lessons are formulated.


\begin{enumerate}

\item (Section~\ref{sec:segtree}) A bottom-up traversal of the {\segtree} has a much simpler
implementation compared to top-down, resulting in a faster execution
of both {\Sum} and {\Update}.

\item (Section~\ref{sec:segtree}) A branch-free implementation of the {\segtree}
is on average $2\times$ faster than a branchy implementation
for all array sizes, $n$, up to $2^{25}$.
This is so because the processor's pipeline is not stalled due to
branches, for a consequent increase in the instruction throughput.
For $n > 2^{25}$, the branchy code
is faster because it saves memory accesses -- the dominant cost
in the runtime for large values of $n$.

In particular, the \emph{combination} of
branchy and branch-free code execution
-- what we called the \emph{two-loop} optimization --
results in a better runtime.

Taking this into account,
we recommend a version of the bottom-up {\segtree}
where an internal node holds the sum of the leaves descending
from its \emph{left} subtree,
because it allows the use of the two-loop optimization
for {\Update} as well (and not only for {\Sum}).

\item (Section~\ref{sec:fentree}) The {\fentree} suffers from cache conflicts for larger values
of $n$, caused by accessing nodes whose memory address is (almost)
a large power of 2. Solving this issue, by offsetting a node from
an original position $i$ to a new position $i + \lfloor i/d \rfloor$
for some $d>0$,
significantly improves the performance of the {\fentree}.

\begin{table}
\centering

\caption{Average speedup factors achieved by the {\fentree}
over the {\segtree}.
\label{tab:speedups_ft}}
\scalebox{1.0}{\begin{tabular}{c c c c}
\toprule

Array Size & \multicolumn{1}{c}{\Sum}
&
& \multicolumn{1}{c}{\Update}
\\

\midrule

\,\,$2^8 < n \leq 2^{16}$ &
1.32$\times$ & & 1.17$\times$ \\

$2^{16} < n \leq 2^{22}$ &
2.43$\times$ & & 1.64$\times$ \\

$2^{22} < n \leq 2^{30}$ &
1.82$\times$ & & 1.50$\times$ \\

\bottomrule
\end{tabular}}
\end{table}

\item (Section~\ref{sec:segment_tree_vs_fenwick_tree}) The {\fentree} is more efficient than
(our optimized version of) the {\segtree} for both queries and updates.
The better efficiency is due to the simplicity of the code
and the less (50\%) average number of loop iterations.
We summarize the speedups achieved by the {\fentree}
over the {\segtree} in Table~\ref{tab:speedups_ft},
for different ranges of $n$.

\item (Section~\ref{sec:segtree_bary}) Although the {\segtree} is outperformed
by the {\fentree}, we can enlarge its branching factor to
a generic quantity $b>2$: this reduces
the height of the {\segtree} for a better cache usage and
enables the use of SIMD instructions.
Such instructions are very effective to
lower the running time of {\Update}, because several values per node
can be updated in parallel.


Compared to the scalar {\Update} algorithm, SIMD can be on average
$2-6\times$ faster depending of the value of $n$.

\item (Section~\ref{sec:segtree_bary}) For the best of performance, we recommend to model the height
of the $b$-ary {\segtree} with a constant known at compile-time, so that the
compiler can execute a specialized code path to handle the specific
value of the height. This completely avoids branches during the execution
of {\Sum} and {\Update}.

The vectorized {\Update} implementation
together with this branch-avoiding optimization
makes the $b$-ary {\segtree} actually faster than the {\fentree}.
We report the speedups achieved by this data structure
over the {\fentree} in Table~\ref{tab:speedups_sst}, for the different
tested combinations of $b$ and $\delta$.
(Recall that $\delta$ represents the bit-width of $\Delta$, the
update value.)

\item (Section~\ref{sec:segtree_bary}) For the most general case where $\delta=64$ bits,
we recommend the use of a $b$-ary {\segtree} with $b=64$.
From Table~\ref{tab:speedups_sst}, we see that this solution offers
an improved trade-off between the running time of {\Sum} and {\Update}:
on average $1.9-5\times$ faster for {\Sum} and up to $1.6\times$ faster
for {\Update} than the {\fentree}.

\item (Section~\ref{sec:segtree_bary}) For the restricted case where $\delta=8$ bits,
we can update even more values in parallel
by \emph{buffering} the updates at each node of the tree.
Considering again Table~\ref{tab:speedups_sst}, for such case
we recommend the use of a $b$-ary {\segtree} with $b=256$.
This solution is faster than a {\fentree} by $1.7-4.7\times$
for {\Sum} and by $1.4-2.5\times$ for {\Update}.


\item (Section~\ref{sec:fentree_bary}) The $b$-ary {\fentree} improves the runtime for queries at the
price of a significant slowdown for updates, compared to the classic
{\fentree} with $b=2$.
This makes the data structure unpractical unless updates are extremely
few compared to queries.

\item (Section~\ref{sec:fentree_bary}) Despite of the larger tree height,
the \emph{blocked} {\fentree}
improves the trade-off of the $b$-ary {\fentree}
(in particular, it improves {\Update} but worsens {\Sum}).
However, it does not beat the classic {\fentree} because the
time spent at each traversed node is much higher
(more cache-misses due to the two-level structure of a node;
more spent cycles due to SIMD).

\item (Section~\ref{sec:fentree_bary}) In order to combine the simplicity of the {\fentree}
with the advantages of blocking $b$ keys together (reduced cache-misses;
SIMD exploitation), a \emph{truncated} {\fentree} can be used.
This data structure improves over the classic {\fentree},
especially for large values of $n$, exposing a trade-off similar
to that of the $b$-ary {\segtree}
(but with the latter being generally better).

\end{enumerate}

\begin{table}
\centering
\caption{Average speedup factors achieved by the $b$-ary {\segtree}
over the {\fentree}.
\label{tab:speedups_sst}}
\subfloat[{\Sum}]{
	\scalebox{1}{\begin{tabular}{c cc c cc}
\toprule

$\delta$
& \multicolumn{2}{c}{64}
&
& \multicolumn{2}{c}{8}
\\

\cmidrule(lr){2-3}
\cmidrule(lr){5-6}

$b$ & 64 & 256 && 64 & 256 \\

\midrule

$\,\,2^8 < n \leq 2^{16}$ &
5.29$\times$ & 6.57$\times$ & & 3.28$\times$ & 4.66$\times$ \\

$2^{16} < n \leq 2^{22}$ &
2.58$\times$ & 3.24$\times$ & & 1.11$\times$ & 1.54$\times$ \\

$2^{22} < n \leq 2^{30}$ &
1.90$\times$ & 2.56$\times$ & & 1.27$\times$ & 1.66$\times$ \\

\bottomrule
\end{tabular}}
}
\subfloat[{\Update}]{
	\scalebox{1}{\begin{tabular}{c cc c cc}
\toprule

$\delta$
& \multicolumn{2}{c}{64}
&
& \multicolumn{2}{c}{8}
\\

\cmidrule(lr){2-3}
\cmidrule(lr){5-6}

$b$ & 64 & 256 && 64 & 256 \\

\midrule

$\,\,2^8 < n \leq 2^{16}$ &
1.62$\times$ & 1.08$\times$ & & 2.08$\times$ & 2.52$\times$ \\

$2^{16} < n \leq 2^{22}$ &
1.16$\times$ & 0.82$\times$ & & 1.56$\times$ & 1.72$\times$ \\

$2^{22} < n \leq 2^{30}$ &
1.05$\times$ & 0.88$\times$ & & 1.35$\times$ & 1.40$\times$ \\

\bottomrule
\end{tabular}}
	\label{tab:speedups_sst_b}
}
\end{table}


Some final remarks follow.
The runtime of {\Update} will improve as SIMD instructions
will become more powerful in future years (e.g., with lower latency),
thus SIMD is a very promising hardware feature that cannot be overlooked
in the design of practical algorithms.

So far we obtained best results using SIMD registers of 128 and 256 bits
(SSE and AVX instruction sets, respectively).
We also made some experiments with the new AVX-512 instruction set
which allows us to use massive load and store instructions comprising
512 bits of memory.
In particular, doubling the size of the used SIMD registers
can be exploited in two different ways:
by either reducing
the number of instructions, or enlarging the branching factor of a node.
We tried both possibilities but did not observe a clear improvement.

A promising avenue for future work would be to consider the \emph{searchable}
version of the problem, i.e., to implement the {\Search} operation.
Note that this operation is actually amenable to SIMD vectorization,
and has parallels with search algorithms in inverted indexes.
With this third operation, the objective would be to use (a specialization of)
the best data structure from this article
-- the $b$-ary {\segtree} with SIMD on updates \emph{and searches} --
to support \emph{rank/select} queries
over mutable bitmaps~\cite{vigna2019}, an important building block
for dynamic succinct data structures.

{The experiments presented in this work were conducted using a single
system configuration, i.e., a specific processor (Intel i9-9940X),
operating system (Linux), and compiler (\textsf{gcc}).
We acknowledge the specificity of our analysis, although it
involves a rather common setup, and
we plan to extend our experimentation to other
configurations as well in future work.
}

\section{ACKNOWLEDGMENTS}
This work was partially supported by the BigDataGrapes (EU H2020 RIA, grant agreement N\textsuperscript{\b{o}}780751), the ``Algorithms, Data Structures and Combinatorics for Machine Learning'' (MIUR-PRIN 2017), and the OK-INSAID (MIUR-PON 2018, grant agreement N\textsuperscript{\b{o}}ARS01\_00917) projects.

\renewcommand{\bibsep}{3.0pt}
\bibliographystyle{ACM-Reference-Format}
\bibliography{bibliography}

\end{document}